\documentclass[floatfix,aps,pra,superscriptaddress,twocolumn,longbibliography]{revtex4-2}
\makeatletter
\renewcommand{\fnum@figure}{Fig. \thefigure}
\usepackage{amsmath}
\usepackage[english]{babel}
\usepackage{bbm}
\usepackage[normalem]{ulem}
\usepackage{times}
\usepackage[normalem]{ulem}
\usepackage{dsfont}
\usepackage{txfonts}
\usepackage{graphicx}
\usepackage{dcolumn}

\usepackage{qcircuit} 
\usepackage{tabularx}
\usepackage{enumerate}
\usepackage[colorlinks=true, citecolor=blue,urlcolor=blue, linkcolor=blue,citebordercolor={1 0 0},linkbordercolor={0 0 1}]{hyperref}
\usepackage{amsfonts,amssymb}
\usepackage{mathrsfs}
\usepackage{subfigure}
\usepackage{pifont}
\usepackage{cleveref}
\usepackage{color}
\usepackage{thmtools}
\usepackage{verbatim}
\usepackage[ruled,vlined]{algorithm2e}
\usepackage{siunitx}
\usepackage{soul}

\newcommand{\mol}[1]{\mathrm{#1}}
\newcommand{\pg}[1]{\rm{#1}}

\usepackage{xr}
\makeatletter
\newcommand*{\addFileDependency}[1]{
  \typeout{(#1)}
  \@addtofilelist{#1}
  \IfFileExists{#1}{}{\typeout{No file #1.}}
}
\makeatother

\newcommand{\sustech}{Department of Physics, Southern University of Science and Technology, Shenzhen 518055, China}
\newcommand{\siqse}{Shenzhen Institute for Quantum Science and Engineering, Southern University of Science and Technology, Shenzhen 518055, China}
\newcommand{\huawei}{Central Research Institute, 2012 Labs, Huawei Technologies}
\newcommand{\tsinghua}{Department of Chemistry, Tsinghua University, Beijing 100084, China}
\newcommand{\sustechchem}{Department of Chemistry, Southern University of Science and Technology, Shenzhen 518055, China}

\begin{document}
\title[]{\texorpdfstring{Toward a Larger Molecular Simulation on the Quantum Computer:\\ Up to 28-Qubit System Accelerated by Point Group Symmetry}{}} 
\author{Changsu Cao}
\affiliation{\huawei}
\affiliation{\tsinghua}
\author{Jiaqi Hu}
\affiliation{\sustech}
\author{Wengang Zhang}
\affiliation{\huawei}
\affiliation{\sustech}
\affiliation{\siqse}

\author{Xusheng Xu}
\affiliation{\huawei}

\author{Dechin Chen}
\affiliation{\huawei}

\author{Fan Yu}
\affiliation{\huawei}

\author{Jun Li}
\affiliation{\tsinghua}
\affiliation{\sustechchem}

\author{Han-Shi Hu}
\email{hshu@mail.tsinghua.edu.cn}
\affiliation{\tsinghua}

\author{Dingshun Lv} \email{ywlds@163.com}
\affiliation{\huawei}
\author{Man-Hong Yung} \email{yung.manhong@huawei.com}
\affiliation{\huawei}
\affiliation{\sustech}
\affiliation{\siqse}

\date{\today}

\begin{abstract}    
The exact evaluation of the molecular ground state in quantum chemistry requires an exponentially increasing computational cost. Quantum computation is a promising way to overcome the exponential problem using polynomial-time quantum algorithms. 
A quantum-classical hybrid optimization scheme known as the variational quantum eigensolver~(VQE) is preferred for noisy intermediate-scale quantum devices.
However, the circuit depth becomes one of the bottlenecks of its application to large molecules of more than 20 qubits. In this work, we employ the point group symmetry to reduce the number of operators in constructing ansatz so as to achieve a more compact quantum circuit.
 We illustrate this methodology with a series of molecules ranging from LiH~(12 qubits) to $\mol{C_2H_4}$~(28 qubits). A significant reduction of up to 82\% of the operator numbers is reached on $\mol{C_2H_4}$, which enables the largest molecule ever numerically simulated by VQE-UCC to the best of our knowledge. This also shed light into the further work of this direction to construct even shallower ansatz with enough expressive power and simulate even larger scale system.
\end{abstract}

\maketitle
\section{Introduction}
 
Quantum computing is proposed to be a promising way to overcome the exponential issue in simulating the energies and properties of the many-electron molecular system by classical computers, as speculated by Feynman in 1982~\cite{feynman1982}. Since then, various quantum algorithms have been developed~\cite{aspuru2005simulated, kassal2008polynomial, Huh2014a, mcardle2020quantum,babbush2018low,zeng2021universal,peruzzo2014variational}, among which, the variational quantum eigensolver (VQE)~\cite{peruzzo2014variational, Yung_2014,endo2020variational,kandala2017hardware,XU2021} is believed to be friendly to near-term quantum devices for its noise-resilient property and a small need for quantum gates, which benefits from its hybrid quantum-classical framework~\cite{o2016scalable}. 

VQE has been applied to simulate chemical systems both experimentally and numerically. The first demonstration of $\mol{H_2}$ (2 qubits) on quantum devices is presented in Ref ~\cite{peruzzo2014variational}. After that, a variety of quantum simulations were performed for  $\mol{BeH_2}$ (6 qubits)~\cite{kandala2017hardware}, $\mol{H_2O}$ (8 qubits)~\cite{Nam2020} and $\mol{H_{12}}$ (12 qubits)~\cite{AIQuantum2020}. 

To benchmark the performance or optimize the algorithm, the numerical results are also presented using virtual quantum simulators on different molecules~\cite{yeter2021benchmarking, lolur2020benchmarking, Kuhn2019}. To date, the largest is 20 qubits for $\mol{H_2O}$ ~\cite{lolur2020benchmarking} with 6-31G basis set.

The scale of the simulation is limited by two correlated factors, the number of controllable qubits and the circuit depth. Although the top record of the controllable qubit number reaches 66~\cite{zuchongzhi}, the depth of quantum circuit is still a problematic limitation on the large scale quantum chemical simulation. 
Therefore, to extend the scale of quantum chemical simulation, a better ansatz initialization requires not only smaller demanding for qubit number but also using less parameters for a more compact quantum circuit.

Unitary coupled-cluster (UCC) ansatz was used when VQE was initially proposed, and has been one of the most popular choices since then. 
In order to represent the molecule with less parameters for a more compact quantum circuit, one feasible method is to screen out the less important parameters base on UCC ansatz by pre-calculations or using adaptive ways~\cite{grimsley2019adaptive,romero2018strategies,fan2021circuit,wecker2015progress}.
Different improved coupled-cluster (CC) ansatz are also proposed and reach hopeful results.~\cite{QCC, Ryabinkin2020,  dallaire2018low, Lee2019}. 

Besides these methods, for the chemical system, using the intrinsic information of it might benefit more to obtain the compact ansatz and introduce less approximations.

For example, the particle number conservation ($U(1)$ symmetry), the fermionic parity conservation~($Z_2$ symmetry), are used to be restrictions in VQE~\cite{bravyi2017tapering, Gard2019, Greene-Diniz2021}. 
The geometric property of molecules described by the point group symmetry will also provide great convenience to the quantum chemical simulation, of which the power has already demonstrated in conventional $ab\ initio$ chemical calculations in classical computers~\cite{CCsym_1,CCsym_2,CCSym3}. 
However, the power of point group symmetry is rarely explored in the quantum computing regime. Currently, the point group symmetry has been used for two purposes: removing qubits in simulation~\cite{setia2020reducing, fischer2019symmetry} and reducing the depth of quantum circuit in quantum computing~\cite{PhysRevA.101.052340, hamiton_sym2}. Setia et al. reduce the qubits in simulation by using the point group to find the permutation matrix to do qubit tapering off~\cite{setia2020reducing}. 
Fischer and Gunlycke map the configuration states based on the point group symmetry to qubits rather than the molecular orbitals, which makes it possible to represent the same molecule with less qubits~\cite{fischer2019symmetry}. 
By applying projectors of symmetry operators to the prepared quantum state, the depth of quantum circuit can be reduced at the cost of more measurements~\cite{PhysRevA.101.052340, hamiton_sym2}.

In this work, we use the point group symmetry to directly reduce the operator numbers in UCC ansatz so that the depth of the quantum circuit is significantly decreased without damage to the accuracy nor expensive auxiliary calculations. 
We present a series of numerical cases with the symmetry reduced unitary coupled-cluster singles and doubles (SymUCCSD) ansatz by the MindQuantum simulator, including $\mol{LiH}$, $\mol{HF}$, $\mol{H_2O}$, $\mol{BeH_2}$, $\mol{CH_4}$, and $\mol{NH_3}$ with its flipping potential energy surface. We have also successfully performed the simulation of 28-qubit $\mol{C_2H_4}$ on the virtual quantum simulator, to the best of our knowledge, the largest size in qubits ever numerically simulated via UCCSD-VQE algorithm. 

\section{Framework}\label{sec:symred}
The variational quantum eigensolver~(VQE) method is originally designed to solve the ground state energies of molecular Hamiltonian~\cite{Yung_2014}. Typically, in these problems, the Hamiltonian under Born-Oppenheimer approximation is usually written in a second-quantized form~\cite{McArdle2018} as 
\begin{equation}\label{secondh}
    	\hat{H}=\sum\limits_{p,q} h_{pq}\hat{a}_p^{\dagger}\hat{a}_q
    	+\frac{1}{2}\sum\limits_{p,q,r,s} g_{pqrs}\hat{a}_p^{\dagger}\hat{a}_r^{\dagger}\hat{a}_{s} \hat{a}_{q},
\end{equation}
where ${\hat{a}_p}^\dag$ and $\hat{a}_p$ denote the fermionic creation operator and annihilation operator associated with $p$-th fermionic mode~(or spin-orbital). The sets of coefficients $\{h_{pq}\}$ and $\{g_{pqrs}\}$ are called one- and two-electron integrals and can be evaluated by classical computers. 
The main idea of VQE is that the parametrized quantum state $\Psi(\Vec{\theta})$ is prepared and measured on a quantum computer, while the parameters are updated in a classical computer following the variational principle $E_0 \leq \min_{\vec{\theta}} {\langle\Psi(\vec{\theta})|\hat{H}|\Psi(\vec{\theta})\rangle}$, where $E_0$ is the ground state energy.

The essential part for VQE is to construct the parametrized quantum state $\Psi(\Vec{\theta})$ that is close enough to the unknown ground state. 
The ansatz derived from the unitary coupled-cluster~(UCC) method~\cite{hoffmann1988unitary,bartlett1989alternative,mcardle2020quantum,shen2017quantum,liu2021variational} is one of the most popular choices.
$\Psi(\Vec{\theta})$ can be constructed as
\begin{equation}
|\Psi(\Vec{\theta})\rangle=e^{\hat{T}(\boldsymbol{\Vec{\theta}})-\hat{T}^{\dagger}(\boldsymbol{\Vec{\theta}})}|\Psi_0\rangle,
\end{equation}
where $\hat{T}$ is the coupled-cluster excitations, and $|\Psi_0\rangle$ is the initial state~(usually a Hartree-Fock state).
The coupled-cluster excitations are usually truncated to single and double excitations, named as UCCSD, 
\begin{gather}
\hat{T}_{1}(\boldsymbol{\theta})=
\sum\limits_{i,a}\hat{t}_{i}^{a}=
\sum\limits_{i,a}\theta_{i}^{a}\hat{a}_a^{\dagger}\hat{a}_i,\\
\hat{T}_{2}(\boldsymbol{\theta})=
\sum\limits_{i,j,a,b}\hat{t}_{ij}^{ab}=
\sum\limits_{i,j,a,b}
\theta_{ij}^{ab}\hat{a}_a^{\dagger}\hat{a}_b^{\dagger}\hat{a}_i \hat{a}_j.
\end{gather}
To implement the VQE circuit of UCC ansatz on a quantum device, the Trotter-Suzuki expansion is needed. Conventionally, the first-order Trotterization is enough to reproduce UCCSD results, while more Trotter steps hardly improve the accuracy but significantly elongate quantum circuit depth.~\cite{o2016scalable, PRA_trott_discuss}
The UCCSD ansatz with first-order Trotterization is expressed as:
	\begin{equation}\label{Trotter}
		|\Psi(\Vec{\theta})\rangle_{Trot} = 
		{{\prod_{i,a}{e^{\hat{t}_{i}^{a}-\hat{t}_{i}^{a\dagger}}}}{\prod_{i,j,a,b}{e^{\hat{t}_{ij}^{ab}-\hat{t}_{ij}^{ab\dagger}}}}}|\Psi_{0}\rangle,
	\end{equation}

For this chemical inspired ansatz, it is natural to utilize the built-in information of the molecule to reduce the computational cost. 
According to the point group, the symmetric properties of the molecular wavefunction can be described by the irreducible representation. The point group symmetry has been employed in classic quantum chemical calculations as a common practice. 
In coupled-cluster~(CC) theory, the amplitude of the excitation operator will vanish unless the corresponding term preserves the totally symmetric irrep, which has been shown by $\mbox{\rm{{\v{C}}{\'a}rsky}}$ et.al.~\cite{CCsym_1} and by Stanon et.al.~\cite{CCsym_2} 
In unitary coupled-cluster theory~(UCC), however, the introduction of the de-excitation operator $\hat{t}^{\dagger}$ prevents natural truncation and the operator terms can not be expressed with finite terms as it is in CC theory. The previous conclusion in CC can not directly apply in UCC and it is not trivial whether the symmetry constraint is still valid. We provide a detailed derivation in the supplemental material that the constraint still holds in SI. Section II. In this sense, only the excitation that belongs to the same irrep of the reference state is valid in UCC ansatz:
\begin{equation}
\forall D\left(e^{\hat{t}-\hat{t}^{\dagger}}\left|\Psi_{0}\right\rangle\right) \neq D\left(\left|\Psi_{0}\right\rangle\right) : e^{\hat{t}-\hat{t}^{\dagger}}=1,
\end{equation}
where $D$ is the irrep of the corresponding wavefunction. 

The key step of our method is to compare the irrep of all the possible excited states with the reference state. The wavefunction is expressed as the Slater-determinant of a collection of molecular orbitals, i.e. $\left|\Psi_{0}\right\rangle=\left|\phi_{1}\overline{\phi}_{1}\phi_{2}\overline{\phi}_{2}...\phi_{n}\overline{\phi}_{n}\right\rangle$, where $\phi$ is the occupied molecular spin-orbital and the bar indicates the different spin. 
The irrep of each molecular orbital can be determined automatically after the HF calculation on classic computers by most of quantum chemical packages like PySCF~\cite{PySCF_1, PySCF_2}. 
Accordingly, the irrep of the excited state (and operator) is determined from the direct product of the irrep of the molecular spin-orbitals by looking up the product table of irreps, which needs no complex numerical computations.  

The overall flow can be summarized in Algorithm~\ref{scheme} and we term this method as SymUCCSD in the following of this work.

\begin{algorithm}[bt] \label{scheme}
    \caption{Scheme of SymUCCSD in VQE} 
    \begin{enumerate}
        \item Initialize the reference state $|\Psi_{0}\rangle$ (Usually the Hartree-Fock ground state). 
        \item \textbf{For} each possible excitation operator $\hat{t}$ in $\hat{T}$ 
        
            \quad \textbf{If} 
            $D\left(e^{\hat{t}-\hat{t}^{\dagger}}\left|\Psi_{0}\right\rangle\right) \neq D\left(\left|\Psi_{0}\right\rangle\right)$
              
              \quad\quad Remove $\hat{t}$ from $\hat{T}$.
              
              \quad \textbf{End}
              
              \textbf{End}
        
        \item  Construct the ansatz operator $e^{\hat{T}-\hat{T}^{\dagger}}$ by the reduced $\hat{T}$. 
        \item Convert the ansatz operator to the quantum circuit and prepare the ansatz. 
        
        \item Perform VQE loop with the generated ansatz until the energy converges or reach the max number of  iterations.
    \end{enumerate}
\end{algorithm}

\section{Numerical results} \label{sec:results}
Using the symmetry reduction method mentioned above, we performed the symmetry reduced UCCSD-VQE on several testing systems, varying from 12 qubits $\mol{LiH}$ to 28 qubits $\mol{C_2H_4}$. Specially, we simulated $\mol{BeH_2}$ under various point group symmetry to study the relationship between the order of the group and the reduction of parameters. The flipping of ammonia, which involves the non-equilibrium geometry structures, was simulated as well. 
All the VQE simulations are performed using the quantum simulator MindQuantum~\cite{mindquantum}. STO-3G basis set was employed in all cases~\cite{sto3g}. The fermion operators are transformed to qubit type operators by Jordan-Wigner transformation~\cite{Jordan1993}. Gradient-based optimization method BFGS is used to minimize the energy expectation value generated by the MindQuantum, of which the convergence threshold is $10^{-6}$. All the geometric structures are obtained from CCCBDB-NIST Database~\cite{NISTCCCBDB}.

\subsection{\texorpdfstring{$\bf{BeH_2}$}{Lg}: under different point group symmetries}

\begin{figure} [hbt] 
    \centering
    \includegraphics[width=0.98\linewidth]{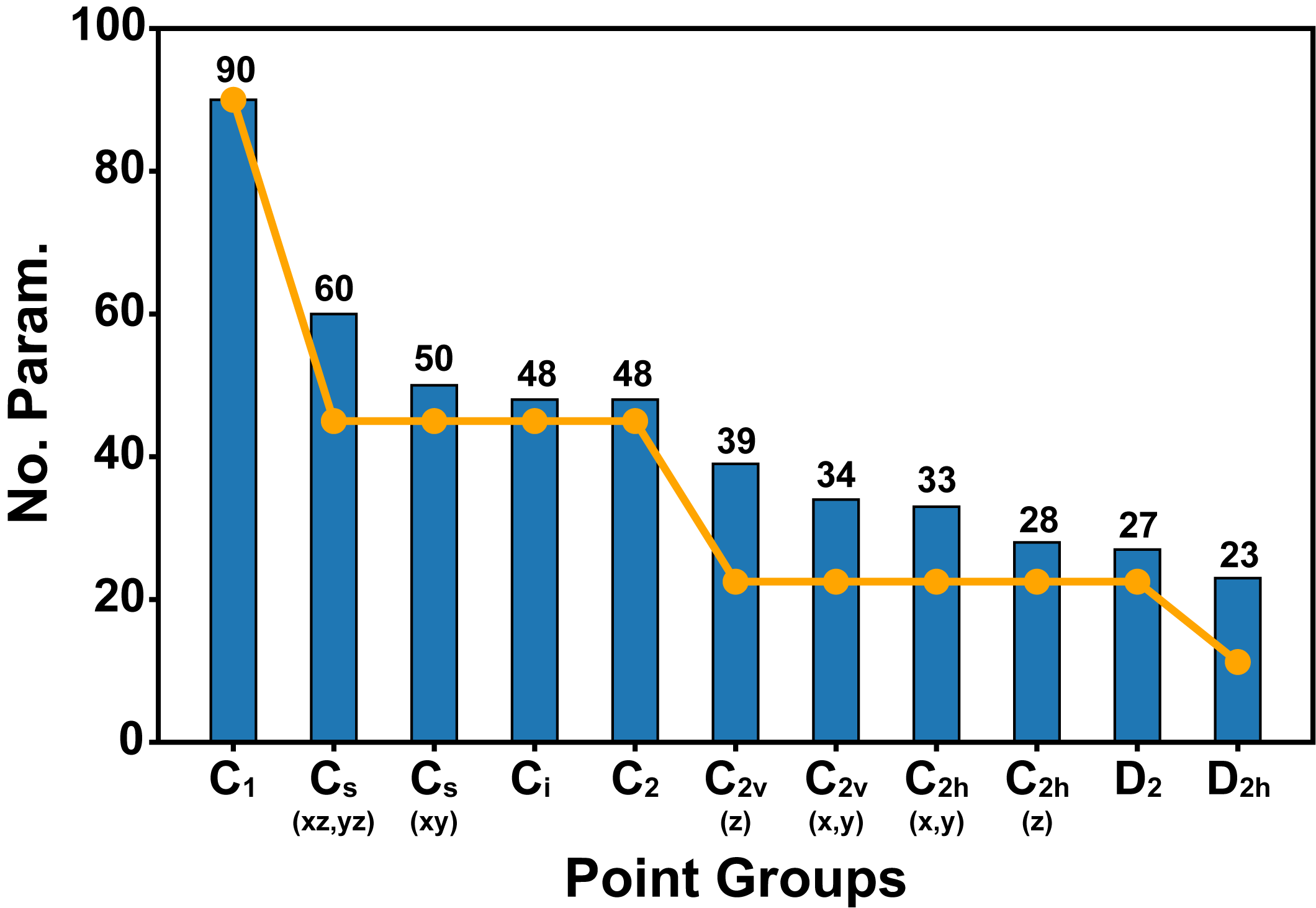}
    \caption{{\bf Parameters reduction of $\bf{BeH_2}$ under various point group.} The orientation of the symmetry elements changes the ratio of reduction slightly. The orange line is the hypothetical situation that the number of configurations belongs to each irrep equal.}
    \label{fig:BeH2_various_sym}
\end{figure}

A specific molecule belongs to different point groups, i.e. the point group with the highest symmetry and its subgroups. Here we simulate $\mol{BeH_2}$ with various point groups assigned to it to investigate the relationship between the order of the group and the reduction of the ansatz. 
$\rm{D_{2h}}$ is the highest-order Abelian point group of $\mol{BeH_2}$ belongs to. Alternatively, the subgroup of $D_{2h}$, i.e. $D_2$, $C_{2h}$, $C_{2v}$, $C_{2}$, $C_{s}$, $C_{i}$ , and $C_{1}$, can also been employed to reduce the ansatz in Section~\ref{sec:symred}. 
We display the number of remained parameters in Fig.~\ref{fig:BeH2_various_sym}, which is approximately proportional to the reciprocal of the order of the group, $1/h$. The reference line of $1/h$ is also depicted.

$\pg{C_{1}}$ group, which is composed of the identity operation exclusively, has only the A irrep and means "no symmetry". Thus SymUCCSD here works exactly as the original UCCSD-VQE calculation. The 2$\rm{^{nd}}$ order groups $\pg{C_s}$, $\pg{C_i}$, and $\pg{C_2}$ include an additional one reflection (mirror plane), inversion (inversion center), or one rotation (rotation axis) operation, respectively. Around half of the parameters are filtered out when employing the 2$\rm{^{nd}}$ order groups while the orientation of symmetry elements will affect the efficiency of the reduction. Similarly, the number of parameters is further reduced when applying the 4$\rm{^{th}}$ order groups, $\pg{C_{2v}}$, $\pg{C_{2h}}$, and $\pg{D_{2}}$, and the 8$\rm{^{th}}$ order group, $\pg{D_{2h}}$. The promotion in parameter reduction becomes small from 4$\rm{^{th}}$ order groups to the 8$\rm{^{th}}$ order group $\pg{D_{2h}}$.

In our method, only parameters belong to referenced irrep would survive. The order of the group equals to the number of irreps, which explains the $1/h$ reduction ratio. The excited terms may distribute unevenly in each irrep especially in small systems. The uneven distribution problem will be improved when it comes to larger systems as we will find below.
\subsection{Simulations of small molecules}

A variety of molecules, i.e. the hydrides containing the 2$\rm{^{ed}}$ row elements, are chosen as the testing cases for benchmark, including $\mol{LiH}$, $\mol{HF}$, $\mol{H_2O}$, $\mol{BeH_2}$, $\mol{NH_3}$, $\mol{CH_4}$.
In SymUCCSD calculations, the highest possible Abelian point group was chosen to simplify the ansatz. Among the testing molecules,
the ratio of the remained parameters ranging from $26\%$ in $\mol{BeH_2}$ with $\pg{D_{2h}}$ to $56\%$ in $\mol{NH_3}$ with $\pg{C_{s}}$. In the worst case that we tested, around half of the parameters are reduced as the lower bound for the symmetric molecular systems, corresponding to the half size of the quantum circuit. 
As shown in Table.~\ref{tab:smallmol}, the energy calculated by SymUCCSD, UCCSD and CCSD results is compared to the reference energies computed by the exact soultion of full configuration interaction~(FCI). The errors are all less than 1.6 mHartree~(chemical accuracy at the level of STO-3G). 
Furthermore, the difference of computed energy between UCCSD without parameter reduction and SymUCCSD is even a few magnitudes smaller. The worst case is no more than 0.005 mHartree, approaching the threshold of computational convergence. The energy difference between UCCSD and SymUCCSD are within the convergence threshold,  indicating that the method actually filter out some redundant parameters without harm to the accuracy.

\begingroup
\squeezetable
\begin{center}
\begin{table*}
\caption {\label{tab:smallmol} 
VQE simulations for small molecules. The simulation scale ranges from 12 to 18 qubits. The original parameters and the parameters after reduction show in the 4$^{th}$ and 5$^{th}$ columns. The number of parameters used of the symmetry reduction ansatz compare to original UCCSD ansatz are shown in the 6$^{th}$ column. The energy differences compared with the FCI energy are shown from 7$^{th}$ to 9$^{th}$ columns with unit in Hartree. The last column display the energy difference between  The equilibrium geometric structures of these molecules are obtained from CCCBDB-NIST Database~\cite{NISTCCCBDB}.
}
\begin{ruledtabular}
\begin{tabular}{ lllllllllll  }
   & Qubits & Sym. & Para.-Before & Para.-After & \%  & $\Delta E_{CCSD}$ &  $\Delta E_{UCCSD}$ & $\Delta E_{SymUCCSD}$ & $ \Delta E_{UCC-SymUCC}$ \\ \hline
$\mol{HF}$ & 12 & $\pg{C_{2v}}$   & 20 & 11 &  55\%  &  \num{2.94e-8}  &   \num{1.82e-5}   & \num{1.38e-5}  &  \num{4.39e-06}   \\
$\mol{LiH}$ & 12 & $\pg{C_{2v}}$ & 44 & 20 &  45\%   &  \num{1.05e-5} &  \num{1.10e-5}    & \num{1.09e-5}   &   \num{1.08e-07}   \\
$\mol{H_2O}$ & 14 & $\pg{C_{2v}}$ & 65 & 26 &  40\%   &  \num{1.17e-4} &  \num{1.19e-4}    & \num{1.09e-4}  &   \num{9.96e-06}  \\
$\mol{BeH_2}$ & 14 & $\pg{D_{2h}}$ & 90 & 23 &  26\%  &  \num{3.94e-4} & \num{3.83e-4}    & \num{3.82e-4}  &   \num{8.89e-07} \\
$\mol{NH_3}$ & 16 & $\pg{C_{s}}$ & 135 & 75 &  56\%   &  \num{2.14e-4} &  \num{1.94e-4}    & \num{1.86e-4}  &   \num{8.44e-06}  \\
$\mol{CH_4}$ & 18 & $\pg{D_{2}}$ & 230 & 65 &  28\%  &  \num{2.30e-4} & \num{2.06e-4}  & \num{1.96e-4}  &   \num{9.94e-06} \\

\end{tabular}
\end{ruledtabular}
\end{table*}
\end{center}
\endgroup

Note that we also performed numerical simulations of ADAPT-VQE~\cite{adaptvqe}, as a comparison and combination with SymUCCSD on $\mol{BeH_2}$. The results show that ADAPT-VQE can efficiently save measurement cost by combining with SymUCCSD, and the operators selected in ADAPT-VQE indeed satisfied the point group symmetry constraint. For further details, we refer reader to SI. Section IV.

To evaluate the performance on real quantum computers, we have compared SymUCCSD with conventional UCCSD on $\mol{H_4}$ with depolarizing noise and limited measurement shots. According to our numerical results in SI. Section V, SymUCCSD is more robust and accurate than UCCSD.

So removing the unfavored operators in the UCCSD ansatz will not sacrifice the accuracy of the result, as supported by SI that the cluster operator should be totally symmetric and not change the symmetry of the reference wavefunction otherwise the generated coupled-cluster wavefunction would not be the eigenvector of the Hamiltonian.

\subsection{Ammonia Flipping}
\begin{figure}[htb]
    \centering
    \includegraphics[width=\linewidth]{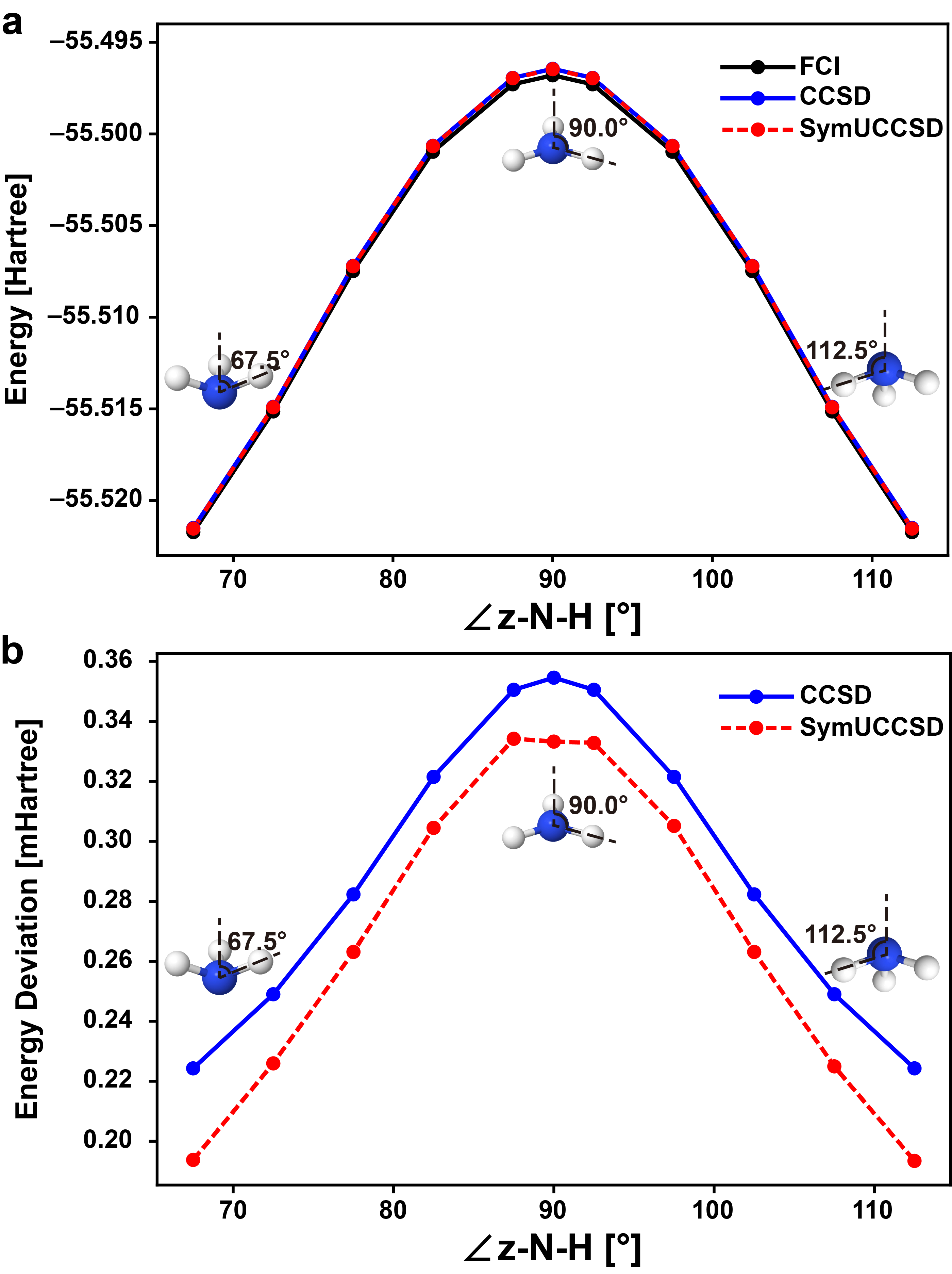} 
    \caption{{\bf The potential energy curve of the ammonia flipping.} {\bf{a}}.Potential energy surface calculated by FCI, CCSD, and the symmetry-reduced UCCSD during ammonia flipping; {\bf{b}}.Energy deviation comparing with FCI energy. The flipping process is described by the angle of the z-axis, N atom and H atom. The unit of the error is mHartree.}
    \label{fig:nh3flip}
\end{figure}

The molecule in the non-equilibrium geometry is important when studying the chemical reaction process which involves bond breaking and conformations changes. To explore the scenario of the non-equilibrium structures, we choose the ammonia flipping process to demonstrate our algorithm. The reaction coordination is computed using PBE density functional approximation~\cite{PBE_DFT} with def2-TZVP~\cite{def2tzvp_basisset} basis set in PySCF~\cite{PySCF_1, PySCF_2}.
The energy profiles regarding FCI, CCSD and SymUCCSD methods are shown in Fig.~\ref{fig:nh3flip}. Concerning the flipping process, the $\pg{C_s}$ symmetry is kept so that the parameter reduction is the same as that in Table~\ref{tab:smallmol}. 

When referring to the FCI results, the error increases when the structure turns from equilibrium to non-equilibrium gemotries and reaches the maximum at the flat structure ($\angle z-N-H=90^{\circ}$), 0.35 mHartree for CCSD and 0.33 mHartree for SymUCCSD. It agrees with our intuition that in the non-equilibrium structure, the multi-reference properties are not negligible anymore. FCI describes the multi-reference properties well, while the truncated CCSD and UCCSD are generally believed to be single-reference methods. The energies calculated by SymUCCSD are slightly lower than those by CCSD. It is probably 
attributed to the introducing of the de-excitation operator in UCC ansatz. 

\subsection{\texorpdfstring{$\bf{C_2H_4}$}{Lg}: Large molecule simulation}

\begin{figure} [htb]
    \centering
    \includegraphics[width=\linewidth]{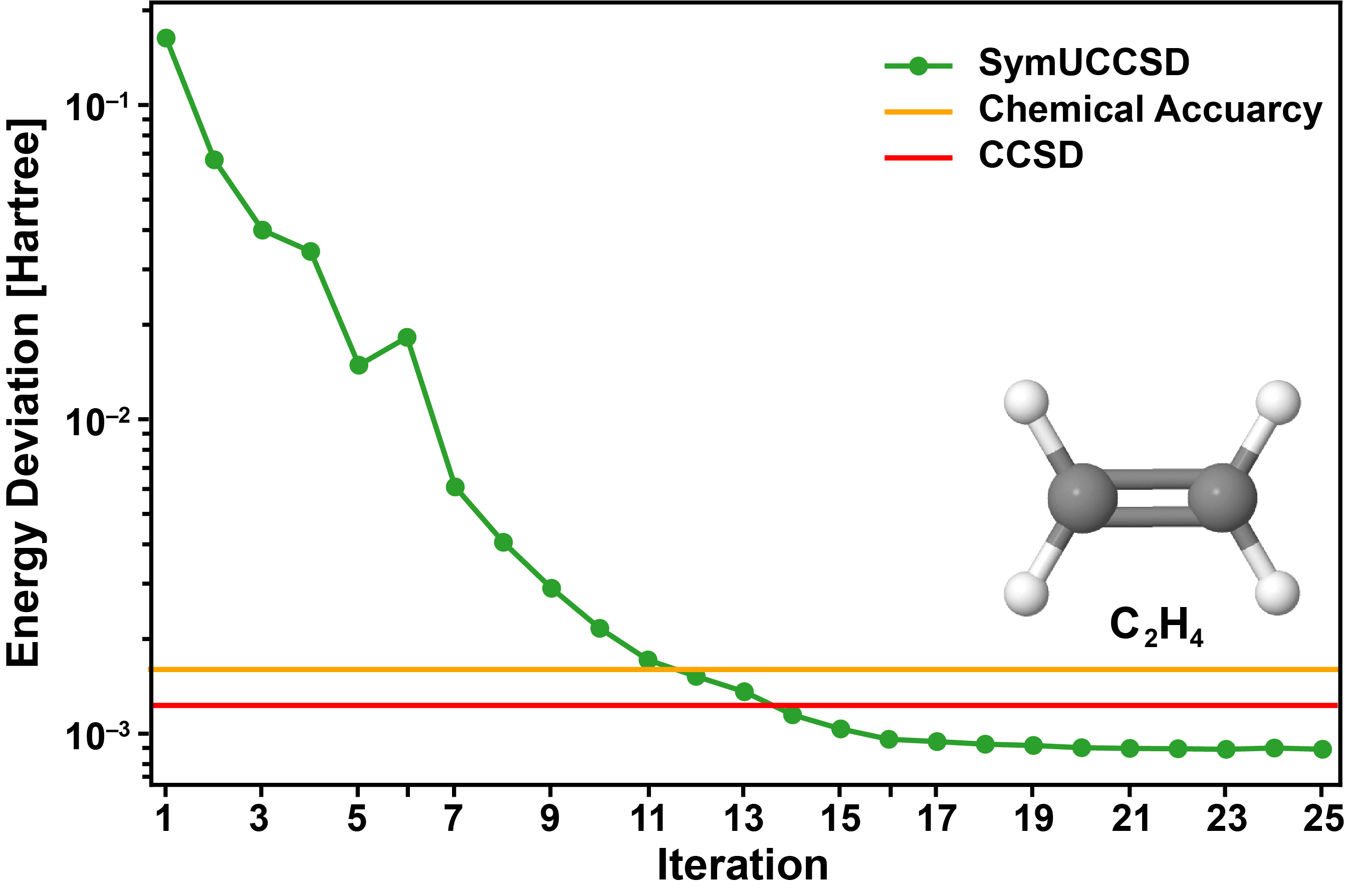}
    \caption{ {\bf The convergence process of SymUCCSD for $\bf{C_2H_4}$}. The green line denotes the energy deviation of SymUCCSD comparing to FCI energy vs. iteration numbers.
    The orange line denotes to energy deviation derivated from CCSD vs. FCI and the green line indicates chemical at the level of STO-3G with reference to FCI results(0.0016 Hartree).}
    \label{fig:c2h4}
\end{figure}

Simulations with more qubits~(indicating more orbitals in the molecule) usually require more parameters and deeper quantum circuits. If we can construct a more compact ansatz, larger molecular simulations will become tractable under the current quantum resource.
To explore the boundary of our method in the current simulator, we present a simulation on 28-qubit $\mol{C_2H_4}$ molecule.
As shown in Fig.~\ref{fig:c2h4}, the calculation converges after 25 iterations and reaches the chemical accuracy at the level of STO-3G around 12$\rm{^{th}}$ iteration. 
The geometric structure of $\mol{C_2H_4}$ belongs to $\pg{D_{2h}}$ point group. The symmetry of the total wavefunction is $\pg{A_g}$ irrep. Table.~\ref{tab:c2h4_irrep} shows that there are 48 single excitations and 1176 double excitations distributing in various irreps. 
Only 18\% of parameters belonging to $\pg{A_g}$ remain after the symmetry reduction, which significantly shortens the depth of the circuit. The greatly reduced number of parameters is an important reason that makes such simulation tractable with the current quantum simulator.

\begingroup
\squeezetable
\begin{center}
\begin{table}
\caption {\label{tab:c2h4_irrep} The number of the excited configurations belongs to each irreps for $\mol{C_2H_4}$.} 

\begin{ruledtabular}
\begin{tabular}{ lll  }
  Irrep. & $T_1$ Num. & $T_2$ Num.  \\ \hline
$A_g$ & 9 & 210 \\
$B_{1g}$ & 8 & 176\\
$B_{2g}$ & 2 & 104\\
$B_{3g}$ & 5 & 110\\
$A_{u}$ & 2 &  104\\
$B_{1u}$ & 3 &  114\\
$B_{2u}$ & 11 & 182\\
$B_{3u}$ & 8 &  176\\
Total &  48    & 1176 \\

\end{tabular}
\end{ruledtabular}
\end{table}
\end{center}
\endgroup

\section{Conclusion}
In this work, we have presented an algorithm to reduce the number of operators needed in UCCSD ansatz by employing the point group symmetry. 
The detailed derivation of this method is attached in Supplemental Material. 
After testing various molecules in different point groups using this method implemented in the MindQuantum simulator~\cite{mindquantum}, we observed that the reduction rate is approximately proportional to $1/h$, where $h$ is the rank of the group. Molecules with higher symmetry such as $\pg{D_{2h}}$ could lead to a larger reduction of the number of operators, meaning more compact quantum circuits. With the help of this scheme, we successfully simulated 28 qubits $\mol{C_2H_4}$ molecule in reasonable computing resource, which is the largest molecule system ever numerically simulated by VQE-UCC to date.We note that with supercomputer we can simulate even larger molecule system. The importance of this large scale simulation also lies in that to this level, the UCCSD ansatz is a good ansatz with enough expressive power\cite{holmes2022connecting, larocca2021theory}. This also shed light into the further work of this direction to construct even shallower ansatz and simulate even larger scale system.

In principle, the point group symmetry is valid and non-exclusive for arbitrary molecular systems using UCC ansatz. 
It is proposed to be compatible with other methods based on excitation operators to compress further the quantum circuit depth, such as low-rank decomposition and low depth circuit~\cite{motta2021low,jastrow2020,  rubin2021compressing,PhysRevXlow_depth,kottmann2021optimize}, (fermion or qubit) ADAPT-VQE or k-UpCCGSD~\cite{fan2021circuit}.
It is proposed to be compatible with other methods based on excitation operators to compress further the quantum circuit depth, such as energy sorting scheme~\cite{fan2021circuit}, (fermion or qubit)~adapt VQE proposed by~\cite{grimsley2019adaptive,tang2021qubit,zhang2020lowdepth} or k-UpCCGSD~\cite{Lee2019}.
Besides such algorithms inspired by the problems to reduce the quantum circuit depth, the quantum circuit compilation~\cite{Trout_2015, jones2020quantum} is also very important and inevitable to implement simulations of the molecule with chemical interests on real quantum hardware, such as superconducting or trapped ion systems.

To further enable even larger scale or more realistic chemistry simulation~\cite{sun2021perturbative}, the current scheme may be incorporate with 
one may treat the current method as a module and incorporate it into the deep VQE method~\cite{fujii2020deep,mizuta2021deep},
the virtual quantum subspace expansion method~\cite{takeshita2020increasing}, the quantum hybrid tensor network~\cite{yuan2021quantum} and the quantum embedding methods~\cite{knizia2012density,rubin2016hybrid,kawashima2021efficient,mineh2021solving,li2021practical,kotliar2006electronic,bauer2016hybrid,rungger2019dynamical}. 

With the advancement of the aforementioned algorithms and the progress in quantum hardware, we anticipate a solid step towards the simulation of realistic molecular systems on the quantum computer soon.

\section*{Acknowledgement}
The authors gratefully thank Jinzhao Sun, Yifei Huang, Weiluo Ren and Junzi Liu for helpful discussions. The calculations were done using KunLun server at Huawei Technologies.

\bibliography{lv.bib}

\begin{thebibliography}{83}%
\makeatletter
\providecommand \@ifxundefined [1]{%
 \@ifx{#1\undefined}
}%
\providecommand \@ifnum [1]{%
 \ifnum #1\expandafter \@firstoftwo
 \else \expandafter \@secondoftwo
 \fi
}%
\providecommand \@ifx [1]{%
 \ifx #1\expandafter \@firstoftwo
 \else \expandafter \@secondoftwo
 \fi
}%
\providecommand \natexlab [1]{#1}%
\providecommand \enquote  [1]{``#1''}%
\providecommand \bibnamefont  [1]{#1}%
\providecommand \bibfnamefont [1]{#1}%
\providecommand \citenamefont [1]{#1}%
\providecommand \href@noop [0]{\@secondoftwo}%
\providecommand \href [0]{\begingroup \@sanitize@url \@href}%
\providecommand \@href[1]{\@@startlink{#1}\@@href}%
\providecommand \@@href[1]{\endgroup#1\@@endlink}%
\providecommand \@sanitize@url [0]{\catcode `\\12\catcode `\$12\catcode
  `\&12\catcode `\#12\catcode `\^12\catcode `\_12\catcode `\%12\relax}%
\providecommand \@@startlink[1]{}%
\providecommand \@@endlink[0]{}%
\providecommand \url  [0]{\begingroup\@sanitize@url \@url }%
\providecommand \@url [1]{\endgroup\@href {#1}{\urlprefix }}%
\providecommand \urlprefix  [0]{URL }%
\providecommand \Eprint [0]{\href }%
\providecommand \doibase [0]{https://doi.org/}%
\providecommand \selectlanguage [0]{\@gobble}%
\providecommand \bibinfo  [0]{\@secondoftwo}%
\providecommand \bibfield  [0]{\@secondoftwo}%
\providecommand \translation [1]{[#1]}%
\providecommand \BibitemOpen [0]{}%
\providecommand \bibitemStop [0]{}%
\providecommand \bibitemNoStop [0]{.\EOS\space}%
\providecommand \EOS [0]{\spacefactor3000\relax}%
\providecommand \BibitemShut  [1]{\csname bibitem#1\endcsname}%
\let\auto@bib@innerbib\@empty
\bibitem [{\citenamefont {Feynman}(1982)}]{feynman1982}%
  \BibitemOpen
  \bibfield  {author} {\bibinfo {author} {\bibfnamefont {R.~P.}\ \bibnamefont
  {Feynman}},\ }\href {https://doi.org/10.1007/BF02650179} {\bibfield
  {journal} {\bibinfo  {journal} {Int. J. Theor. Phys.}\ }\textbf {\bibinfo
  {volume} {21}},\ \bibinfo {pages} {467} (\bibinfo {year} {1982})}\BibitemShut
  {NoStop}%
\bibitem [{\citenamefont {Aspuru-Guzik}\ \emph {et~al.}(2005)\citenamefont
  {Aspuru-Guzik}, \citenamefont {Dutoi}, \citenamefont {Love},\ and\
  \citenamefont {Head-Gordon}}]{aspuru2005simulated}%
  \BibitemOpen
  \bibfield  {author} {\bibinfo {author} {\bibfnamefont {A.}~\bibnamefont
  {Aspuru-Guzik}}, \bibinfo {author} {\bibfnamefont {A.~D.}\ \bibnamefont
  {Dutoi}}, \bibinfo {author} {\bibfnamefont {P.~J.}\ \bibnamefont {Love}},\
  and\ \bibinfo {author} {\bibfnamefont {M.}~\bibnamefont {Head-Gordon}},\
  }\href {https://doi.org/10.1126/science.1113479} {\bibfield  {journal}
  {\bibinfo  {journal} {Science}\ }\textbf {\bibinfo {volume} {309}},\ \bibinfo
  {pages} {1704} (\bibinfo {year} {2005})}\BibitemShut {NoStop}%
\bibitem [{\citenamefont {Kassal}\ \emph {et~al.}(2008)\citenamefont {Kassal},
  \citenamefont {Jordan}, \citenamefont {Love}, \citenamefont {Mohseni},\ and\
  \citenamefont {Aspuru-Guzik}}]{kassal2008polynomial}%
  \BibitemOpen
  \bibfield  {author} {\bibinfo {author} {\bibfnamefont {I.}~\bibnamefont
  {Kassal}}, \bibinfo {author} {\bibfnamefont {S.~P.}\ \bibnamefont {Jordan}},
  \bibinfo {author} {\bibfnamefont {P.~J.}\ \bibnamefont {Love}}, \bibinfo
  {author} {\bibfnamefont {M.}~\bibnamefont {Mohseni}},\ and\ \bibinfo {author}
  {\bibfnamefont {A.}~\bibnamefont {Aspuru-Guzik}},\ }\href
  {https://doi.org/10.1073/pnas.0808245105} {\bibfield  {journal} {\bibinfo
  {journal} {PNAS}\ }\textbf {\bibinfo {volume} {105}},\ \bibinfo {pages}
  {18681} (\bibinfo {year} {2008})}\BibitemShut {NoStop}%
\bibitem [{\citenamefont {Huh}\ \emph {et~al.}(2015)\citenamefont {Huh},
  \citenamefont {Guerreschi}, \citenamefont {Peropadre}, \citenamefont
  {Mcclean},\ and\ \citenamefont {Aspuru-Guzik}}]{Huh2014a}%
  \BibitemOpen
  \bibfield  {author} {\bibinfo {author} {\bibfnamefont {J.}~\bibnamefont
  {Huh}}, \bibinfo {author} {\bibfnamefont {G.~G.}\ \bibnamefont {Guerreschi}},
  \bibinfo {author} {\bibfnamefont {B.}~\bibnamefont {Peropadre}}, \bibinfo
  {author} {\bibfnamefont {J.~R.}\ \bibnamefont {Mcclean}},\ and\ \bibinfo
  {author} {\bibfnamefont {A.}~\bibnamefont {Aspuru-Guzik}},\ }\href
  {https://doi.org/10.1038/nphoton.2015.153} {\bibfield  {journal} {\bibinfo
  {journal} {Nat. Photonics}\ }\textbf {\bibinfo {volume} {9}},\ \bibinfo
  {pages} {615} (\bibinfo {year} {2015})}\BibitemShut {NoStop}%
\bibitem [{\citenamefont {McArdle}\ \emph
  {et~al.}(2020{\natexlab{a}})\citenamefont {McArdle}, \citenamefont {Endo},
  \citenamefont {Aspuru-Guzik}, \citenamefont {Benjamin},\ and\ \citenamefont
  {Yuan}}]{mcardle2020quantum}%
  \BibitemOpen
  \bibfield  {author} {\bibinfo {author} {\bibfnamefont {S.}~\bibnamefont
  {McArdle}}, \bibinfo {author} {\bibfnamefont {S.}~\bibnamefont {Endo}},
  \bibinfo {author} {\bibfnamefont {A.}~\bibnamefont {Aspuru-Guzik}}, \bibinfo
  {author} {\bibfnamefont {S.~C.}\ \bibnamefont {Benjamin}},\ and\ \bibinfo
  {author} {\bibfnamefont {X.}~\bibnamefont {Yuan}},\ }\href
  {https://doi.org/10.1103/RevModPhys.92.015003} {\bibfield  {journal}
  {\bibinfo  {journal} {Rev. Mod. Phys.}\ }\textbf {\bibinfo {volume} {92}},\
  \bibinfo {pages} {015003} (\bibinfo {year} {2020}{\natexlab{a}})}\BibitemShut
  {NoStop}%
\bibitem [{\citenamefont {Babbush}\ \emph
  {et~al.}(2018{\natexlab{a}})\citenamefont {Babbush}, \citenamefont {Wiebe},
  \citenamefont {McClean}, \citenamefont {McClain}, \citenamefont {Neven},\
  and\ \citenamefont {Chan}}]{babbush2018low}%
  \BibitemOpen
  \bibfield  {author} {\bibinfo {author} {\bibfnamefont {R.}~\bibnamefont
  {Babbush}}, \bibinfo {author} {\bibfnamefont {N.}~\bibnamefont {Wiebe}},
  \bibinfo {author} {\bibfnamefont {J.}~\bibnamefont {McClean}}, \bibinfo
  {author} {\bibfnamefont {J.}~\bibnamefont {McClain}}, \bibinfo {author}
  {\bibfnamefont {H.}~\bibnamefont {Neven}},\ and\ \bibinfo {author}
  {\bibfnamefont {G.~K.-L.}\ \bibnamefont {Chan}},\ }\href
  {https://doi.org/10.1103/PhysRevX.8.011044} {\bibfield  {journal} {\bibinfo
  {journal} {Phys. Rev. X}\ }\textbf {\bibinfo {volume} {8}},\ \bibinfo {pages}
  {011044} (\bibinfo {year} {2018}{\natexlab{a}})}\BibitemShut {NoStop}%
\bibitem [{\citenamefont {Zeng}\ \emph {et~al.}(2021)\citenamefont {Zeng},
  \citenamefont {Sun},\ and\ \citenamefont {Yuan}}]{zeng2021universal}%
  \BibitemOpen
  \bibfield  {author} {\bibinfo {author} {\bibfnamefont {P.}~\bibnamefont
  {Zeng}}, \bibinfo {author} {\bibfnamefont {J.}~\bibnamefont {Sun}},\ and\
  \bibinfo {author} {\bibfnamefont {X.}~\bibnamefont {Yuan}},\ }\href@noop {}
  {\bibinfo {title} {Universal quantum algorithmic cooling on a quantum
  computer}} (\bibinfo {year} {2021}),\ \Eprint
  {https://arxiv.org/abs/2109.15304} {arXiv:2109.15304} \BibitemShut {NoStop}%
\bibitem [{\citenamefont {Peruzzo}\ \emph {et~al.}(2014)\citenamefont
  {Peruzzo}, \citenamefont {McClean}, \citenamefont {Shadbolt}, \citenamefont
  {Yung}, \citenamefont {Zhou}, \citenamefont {Love}, \citenamefont
  {Aspuru-Guzik},\ and\ \citenamefont {O'Brien}}]{peruzzo2014variational}%
  \BibitemOpen
  \bibfield  {author} {\bibinfo {author} {\bibfnamefont {A.}~\bibnamefont
  {Peruzzo}}, \bibinfo {author} {\bibfnamefont {J.}~\bibnamefont {McClean}},
  \bibinfo {author} {\bibfnamefont {P.}~\bibnamefont {Shadbolt}}, \bibinfo
  {author} {\bibfnamefont {M.-H.}\ \bibnamefont {Yung}}, \bibinfo {author}
  {\bibfnamefont {X.-Q.}\ \bibnamefont {Zhou}}, \bibinfo {author}
  {\bibfnamefont {P.~J.}\ \bibnamefont {Love}}, \bibinfo {author}
  {\bibfnamefont {A.}~\bibnamefont {Aspuru-Guzik}},\ and\ \bibinfo {author}
  {\bibfnamefont {J.~L.}\ \bibnamefont {O'Brien}},\ }\href
  {https://doi.org/10.1038/ncomms5213} {\bibfield  {journal} {\bibinfo
  {journal} {Nat. Commun.}\ }\textbf {\bibinfo {volume} {5}},\ \bibinfo {pages}
  {4213} (\bibinfo {year} {2014})}\BibitemShut {NoStop}%
\bibitem [{\citenamefont {Yung}\ \emph {et~al.}(2014)\citenamefont {Yung},
  \citenamefont {Casanova}, \citenamefont {Mezzacapo}, \citenamefont {McClean},
  \citenamefont {Lamata}, \citenamefont {Aspuru-Guzik},\ and\ \citenamefont
  {Solano}}]{Yung_2014}%
  \BibitemOpen
  \bibfield  {author} {\bibinfo {author} {\bibfnamefont {M.-H.}\ \bibnamefont
  {Yung}}, \bibinfo {author} {\bibfnamefont {J.}~\bibnamefont {Casanova}},
  \bibinfo {author} {\bibfnamefont {A.}~\bibnamefont {Mezzacapo}}, \bibinfo
  {author} {\bibfnamefont {J.}~\bibnamefont {McClean}}, \bibinfo {author}
  {\bibfnamefont {L.}~\bibnamefont {Lamata}}, \bibinfo {author} {\bibfnamefont
  {A.}~\bibnamefont {Aspuru-Guzik}},\ and\ \bibinfo {author} {\bibfnamefont
  {E.}~\bibnamefont {Solano}},\ }\href {https://doi.org/10.1038/srep03589}
  {\bibfield  {journal} {\bibinfo  {journal} {Sci. Rep.}\ }\textbf {\bibinfo
  {volume} {4}},\ \bibinfo {pages} {3589} (\bibinfo {year} {2014})}\BibitemShut
  {NoStop}%
\bibitem [{\citenamefont {Endo}\ \emph {et~al.}(2020)\citenamefont {Endo},
  \citenamefont {Sun}, \citenamefont {Li}, \citenamefont {Benjamin},\ and\
  \citenamefont {Yuan}}]{endo2020variational}%
  \BibitemOpen
  \bibfield  {author} {\bibinfo {author} {\bibfnamefont {S.}~\bibnamefont
  {Endo}}, \bibinfo {author} {\bibfnamefont {J.}~\bibnamefont {Sun}}, \bibinfo
  {author} {\bibfnamefont {Y.}~\bibnamefont {Li}}, \bibinfo {author}
  {\bibfnamefont {S.~C.}\ \bibnamefont {Benjamin}},\ and\ \bibinfo {author}
  {\bibfnamefont {X.}~\bibnamefont {Yuan}},\ }\href
  {https://doi.org/10.1103/PhysRevLett.125.010501} {\bibfield  {journal}
  {\bibinfo  {journal} {Phys. Rev. Lett.}\ }\textbf {\bibinfo {volume} {125}},\
  \bibinfo {pages} {010501} (\bibinfo {year} {2020})}\BibitemShut {NoStop}%
\bibitem [{\citenamefont {Kandala}\ \emph {et~al.}(2017)\citenamefont
  {Kandala}, \citenamefont {Mezzacapo}, \citenamefont {Temme}, \citenamefont
  {Takita}, \citenamefont {Brink}, \citenamefont {Chow},\ and\ \citenamefont
  {Gambetta}}]{kandala2017hardware}%
  \BibitemOpen
  \bibfield  {author} {\bibinfo {author} {\bibfnamefont {A.}~\bibnamefont
  {Kandala}}, \bibinfo {author} {\bibfnamefont {A.}~\bibnamefont {Mezzacapo}},
  \bibinfo {author} {\bibfnamefont {K.}~\bibnamefont {Temme}}, \bibinfo
  {author} {\bibfnamefont {M.}~\bibnamefont {Takita}}, \bibinfo {author}
  {\bibfnamefont {M.}~\bibnamefont {Brink}}, \bibinfo {author} {\bibfnamefont
  {J.~M.}\ \bibnamefont {Chow}},\ and\ \bibinfo {author} {\bibfnamefont
  {J.~M.}\ \bibnamefont {Gambetta}},\ }\href
  {https://doi.org/10.1038/nature23879} {\bibfield  {journal} {\bibinfo
  {journal} {Nature}\ }\textbf {\bibinfo {volume} {549}},\ \bibinfo {pages}
  {242} (\bibinfo {year} {2017})}\BibitemShut {NoStop}%
\bibitem [{\citenamefont {Xu}\ \emph {et~al.}(2021)\citenamefont {Xu},
  \citenamefont {Sun}, \citenamefont {Endo}, \citenamefont {Li}, \citenamefont
  {Benjamin},\ and\ \citenamefont {Yuan}}]{XU2021}%
  \BibitemOpen
  \bibfield  {author} {\bibinfo {author} {\bibfnamefont {X.}~\bibnamefont
  {Xu}}, \bibinfo {author} {\bibfnamefont {J.}~\bibnamefont {Sun}}, \bibinfo
  {author} {\bibfnamefont {S.}~\bibnamefont {Endo}}, \bibinfo {author}
  {\bibfnamefont {Y.}~\bibnamefont {Li}}, \bibinfo {author} {\bibfnamefont
  {S.~C.}\ \bibnamefont {Benjamin}},\ and\ \bibinfo {author} {\bibfnamefont
  {X.}~\bibnamefont {Yuan}},\ }\href
  {https://doi.org/https://doi.org/10.1016/j.scib.2021.06.023} {\bibfield
  {journal} {\bibinfo  {journal} {Sci. Bull.}\ }\textbf {\bibinfo {volume}
  {66}},\ \bibinfo {pages} {2181} (\bibinfo {year} {2021})}\BibitemShut
  {NoStop}%
\bibitem [{\citenamefont {O'Malley}\ \emph {et~al.}(2016)\citenamefont
  {O'Malley}, \citenamefont {Babbush}, \citenamefont {Kivlichan}, \citenamefont
  {Romero}, \citenamefont {McClean}, \citenamefont {Barends}, \citenamefont
  {Kelly}, \citenamefont {Roushan}, \citenamefont {Tranter}, \citenamefont
  {Ding} \emph {et~al.}}]{o2016scalable}%
  \BibitemOpen
  \bibfield  {author} {\bibinfo {author} {\bibfnamefont {P.~J.~J.}\
  \bibnamefont {O'Malley}}, \bibinfo {author} {\bibfnamefont {R.}~\bibnamefont
  {Babbush}}, \bibinfo {author} {\bibfnamefont {I.~D.}\ \bibnamefont
  {Kivlichan}}, \bibinfo {author} {\bibfnamefont {J.}~\bibnamefont {Romero}},
  \bibinfo {author} {\bibfnamefont {J.~R.}\ \bibnamefont {McClean}}, \bibinfo
  {author} {\bibfnamefont {R.}~\bibnamefont {Barends}}, \bibinfo {author}
  {\bibfnamefont {J.}~\bibnamefont {Kelly}}, \bibinfo {author} {\bibfnamefont
  {P.}~\bibnamefont {Roushan}}, \bibinfo {author} {\bibfnamefont
  {A.}~\bibnamefont {Tranter}}, \bibinfo {author} {\bibfnamefont
  {N.}~\bibnamefont {Ding}}, \emph {et~al.},\ }\href
  {https://doi.org/10.1103/PhysRevX.6.031007} {\bibfield  {journal} {\bibinfo
  {journal} {Phys. Rev. X}\ }\textbf {\bibinfo {volume} {6}},\ \bibinfo {pages}
  {031007} (\bibinfo {year} {2016})}\BibitemShut {NoStop}%
\bibitem [{\citenamefont {Nam}\ \emph {et~al.}(2020)\citenamefont {Nam},
  \citenamefont {Chen}, \citenamefont {Pisenti}, \citenamefont {Wright},
  \citenamefont {Delaney}, \citenamefont {Maslov}, \citenamefont {Brown},
  \citenamefont {Allen}, \citenamefont {Amini}, \citenamefont {Apisdorf} \emph
  {et~al.}}]{Nam2020}%
  \BibitemOpen
  \bibfield  {author} {\bibinfo {author} {\bibfnamefont {Y.}~\bibnamefont
  {Nam}}, \bibinfo {author} {\bibfnamefont {J.-S.}\ \bibnamefont {Chen}},
  \bibinfo {author} {\bibfnamefont {N.~C.}\ \bibnamefont {Pisenti}}, \bibinfo
  {author} {\bibfnamefont {K.}~\bibnamefont {Wright}}, \bibinfo {author}
  {\bibfnamefont {C.}~\bibnamefont {Delaney}}, \bibinfo {author} {\bibfnamefont
  {D.}~\bibnamefont {Maslov}}, \bibinfo {author} {\bibfnamefont {K.~R.}\
  \bibnamefont {Brown}}, \bibinfo {author} {\bibfnamefont {S.}~\bibnamefont
  {Allen}}, \bibinfo {author} {\bibfnamefont {J.~M.}\ \bibnamefont {Amini}},
  \bibinfo {author} {\bibfnamefont {J.}~\bibnamefont {Apisdorf}}, \emph
  {et~al.},\ }\href {https://doi.org/10.1038/s41534-020-0259-3} {\bibfield
  {journal} {\bibinfo  {journal} {npj Quantum Inf.}\ }\textbf {\bibinfo
  {volume} {6}},\ \bibinfo {pages} {33} (\bibinfo {year} {2020})}\BibitemShut
  {NoStop}%
\bibitem [{\citenamefont {Arute}\ \emph {et~al.}(2020)\citenamefont {Arute},
  \citenamefont {Arya}, \citenamefont {Babbush}, \citenamefont {Bacon},
  \citenamefont {Bardin}, \citenamefont {Barends}, \citenamefont {Boixo},
  \citenamefont {Broughton}, \citenamefont {Buckley}, \citenamefont {Buell}
  \emph {et~al.}}]{AIQuantum2020}%
  \BibitemOpen
  \bibfield  {author} {\bibinfo {author} {\bibfnamefont {F.}~\bibnamefont
  {Arute}}, \bibinfo {author} {\bibfnamefont {K.}~\bibnamefont {Arya}},
  \bibinfo {author} {\bibfnamefont {R.}~\bibnamefont {Babbush}}, \bibinfo
  {author} {\bibfnamefont {D.}~\bibnamefont {Bacon}}, \bibinfo {author}
  {\bibfnamefont {J.~C.}\ \bibnamefont {Bardin}}, \bibinfo {author}
  {\bibfnamefont {R.}~\bibnamefont {Barends}}, \bibinfo {author} {\bibfnamefont
  {S.}~\bibnamefont {Boixo}}, \bibinfo {author} {\bibfnamefont
  {M.}~\bibnamefont {Broughton}}, \bibinfo {author} {\bibfnamefont {B.~B.}\
  \bibnamefont {Buckley}}, \bibinfo {author} {\bibfnamefont {D.~A.}\
  \bibnamefont {Buell}}, \emph {et~al.},\ }\href
  {https://doi.org/10.1126/science.abb9811} {\bibfield  {journal} {\bibinfo
  {journal} {Science}\ }\textbf {\bibinfo {volume} {369}},\ \bibinfo {pages}
  {1084} (\bibinfo {year} {2020})}\BibitemShut {NoStop}%
\bibitem [{\citenamefont {Yeter-Aydeniz}\ \emph {et~al.}(2021)\citenamefont
  {Yeter-Aydeniz}, \citenamefont {Gard}, \citenamefont {Jakowski},
  \citenamefont {Majumder}, \citenamefont {Barron}, \citenamefont {Siopsis},
  \citenamefont {Humble},\ and\ \citenamefont
  {Pooser}}]{yeter2021benchmarking}%
  \BibitemOpen
  \bibfield  {author} {\bibinfo {author} {\bibfnamefont {K.}~\bibnamefont
  {Yeter-Aydeniz}}, \bibinfo {author} {\bibfnamefont {B.~T.}\ \bibnamefont
  {Gard}}, \bibinfo {author} {\bibfnamefont {J.}~\bibnamefont {Jakowski}},
  \bibinfo {author} {\bibfnamefont {S.}~\bibnamefont {Majumder}}, \bibinfo
  {author} {\bibfnamefont {G.~S.}\ \bibnamefont {Barron}}, \bibinfo {author}
  {\bibfnamefont {G.}~\bibnamefont {Siopsis}}, \bibinfo {author} {\bibfnamefont
  {T.~S.}\ \bibnamefont {Humble}},\ and\ \bibinfo {author} {\bibfnamefont
  {R.~C.}\ \bibnamefont {Pooser}},\ }\href
  {https://doi.org/https://doi.org/10.1002/qute.202100012} {\bibfield
  {journal} {\bibinfo  {journal} {Adv. Quantum Technol.}\ }\textbf {\bibinfo
  {volume} {4}},\ \bibinfo {pages} {2100012} (\bibinfo {year}
  {2021})}\BibitemShut {NoStop}%
\bibitem [{\citenamefont {Lolur}\ \emph {et~al.}(2020)\citenamefont {Lolur},
  \citenamefont {Rahm}, \citenamefont {Skogh}, \citenamefont
  {Garc{\'\i}a-{\'A}lvarez},\ and\ \citenamefont
  {Wendin}}]{lolur2020benchmarking}%
  \BibitemOpen
  \bibfield  {author} {\bibinfo {author} {\bibfnamefont {P.}~\bibnamefont
  {Lolur}}, \bibinfo {author} {\bibfnamefont {M.}~\bibnamefont {Rahm}},
  \bibinfo {author} {\bibfnamefont {M.}~\bibnamefont {Skogh}}, \bibinfo
  {author} {\bibfnamefont {L.}~\bibnamefont {Garc{\'\i}a-{\'A}lvarez}},\ and\
  \bibinfo {author} {\bibfnamefont {G.}~\bibnamefont {Wendin}},\ }\href@noop {}
  {\bibinfo {title} {Benchmarking the variational quantum eigensolver through
  simulation of the ground state energy of prebiotic molecules on
  high-performance computers}} (\bibinfo {year} {2020}),\ \Eprint
  {https://arxiv.org/abs/2010.13578} {arXiv:2010.13578} \BibitemShut {NoStop}%
\bibitem [{\citenamefont {K{\"{u}}hn}\ \emph {et~al.}(2019)\citenamefont
  {K{\"{u}}hn}, \citenamefont {Zanker}, \citenamefont {Deglmann}, \citenamefont
  {Marthaler},\ and\ \citenamefont {Wei{\ss}}}]{Kuhn2019}%
  \BibitemOpen
  \bibfield  {author} {\bibinfo {author} {\bibfnamefont {M.}~\bibnamefont
  {K{\"{u}}hn}}, \bibinfo {author} {\bibfnamefont {S.}~\bibnamefont {Zanker}},
  \bibinfo {author} {\bibfnamefont {P.}~\bibnamefont {Deglmann}}, \bibinfo
  {author} {\bibfnamefont {M.}~\bibnamefont {Marthaler}},\ and\ \bibinfo
  {author} {\bibfnamefont {H.}~\bibnamefont {Wei{\ss}}},\ }\href
  {https://doi.org/10.1021/acs.jctc.9b00236} {\bibfield  {journal} {\bibinfo
  {journal} {J. Chem. Theory Comput.}\ }\textbf {\bibinfo {volume} {15}},\
  \bibinfo {pages} {4764} (\bibinfo {year} {2019})}\BibitemShut {NoStop}%
\bibitem [{\citenamefont {Wu}\ \emph {et~al.}(2021)\citenamefont {Wu},
  \citenamefont {Bao}, \citenamefont {Cao}, \citenamefont {Chen}, \citenamefont
  {Chen}, \citenamefont {Chen}, \citenamefont {Chung}, \citenamefont {Deng},
  \citenamefont {Du}, \citenamefont {Fan} \emph {et~al.}}]{zuchongzhi}%
  \BibitemOpen
  \bibfield  {author} {\bibinfo {author} {\bibfnamefont {Y.}~\bibnamefont
  {Wu}}, \bibinfo {author} {\bibfnamefont {W.-S.}\ \bibnamefont {Bao}},
  \bibinfo {author} {\bibfnamefont {S.}~\bibnamefont {Cao}}, \bibinfo {author}
  {\bibfnamefont {F.}~\bibnamefont {Chen}}, \bibinfo {author} {\bibfnamefont
  {M.-C.}\ \bibnamefont {Chen}}, \bibinfo {author} {\bibfnamefont
  {X.}~\bibnamefont {Chen}}, \bibinfo {author} {\bibfnamefont {T.-H.}\
  \bibnamefont {Chung}}, \bibinfo {author} {\bibfnamefont {H.}~\bibnamefont
  {Deng}}, \bibinfo {author} {\bibfnamefont {Y.}~\bibnamefont {Du}}, \bibinfo
  {author} {\bibfnamefont {D.}~\bibnamefont {Fan}}, \emph {et~al.},\
  }\href@noop {} {\bibinfo {title} {Strong quantum computational advantage
  using a superconducting quantum processor}} (\bibinfo {year}
  {2021})\BibitemShut {NoStop}%
\bibitem [{\citenamefont {Grimsley}\ \emph
  {et~al.}(2019{\natexlab{a}})\citenamefont {Grimsley}, \citenamefont
  {Economou}, \citenamefont {Barnes},\ and\ \citenamefont
  {Mayhall}}]{grimsley2019adaptive}%
  \BibitemOpen
  \bibfield  {author} {\bibinfo {author} {\bibfnamefont {H.~R.}\ \bibnamefont
  {Grimsley}}, \bibinfo {author} {\bibfnamefont {S.~E.}\ \bibnamefont
  {Economou}}, \bibinfo {author} {\bibfnamefont {E.}~\bibnamefont {Barnes}},\
  and\ \bibinfo {author} {\bibfnamefont {N.~J.}\ \bibnamefont {Mayhall}},\
  }\href {https://doi.org/10.1038/s41467-019-10988-2} {\bibfield  {journal}
  {\bibinfo  {journal} {Nat. Commun.}\ }\textbf {\bibinfo {volume} {10}},\
  \bibinfo {pages} {3007} (\bibinfo {year} {2019}{\natexlab{a}})}\BibitemShut
  {NoStop}%
\bibitem [{\citenamefont {Romero}\ \emph {et~al.}(2018)\citenamefont {Romero},
  \citenamefont {Babbush}, \citenamefont {McClean}, \citenamefont {Hempel},
  \citenamefont {Love},\ and\ \citenamefont
  {Aspuru-Guzik}}]{romero2018strategies}%
  \BibitemOpen
  \bibfield  {author} {\bibinfo {author} {\bibfnamefont {J.}~\bibnamefont
  {Romero}}, \bibinfo {author} {\bibfnamefont {R.}~\bibnamefont {Babbush}},
  \bibinfo {author} {\bibfnamefont {J.~R.}\ \bibnamefont {McClean}}, \bibinfo
  {author} {\bibfnamefont {C.}~\bibnamefont {Hempel}}, \bibinfo {author}
  {\bibfnamefont {P.~J.}\ \bibnamefont {Love}},\ and\ \bibinfo {author}
  {\bibfnamefont {A.}~\bibnamefont {Aspuru-Guzik}},\ }\href
  {https://doi.org/10.1088/2058-9565/aad3e4} {\bibfield  {journal} {\bibinfo
  {journal} {Quantum Sci. Technol.}\ }\textbf {\bibinfo {volume} {4}},\
  \bibinfo {pages} {014008} (\bibinfo {year} {2018})}\BibitemShut {NoStop}%
\bibitem [{\citenamefont {Fan}\ \emph {et~al.}(2021)\citenamefont {Fan},
  \citenamefont {Cao}, \citenamefont {Xu}, \citenamefont {Li}, \citenamefont
  {Lv},\ and\ \citenamefont {Yung}}]{fan2021circuit}%
  \BibitemOpen
  \bibfield  {author} {\bibinfo {author} {\bibfnamefont {Y.}~\bibnamefont
  {Fan}}, \bibinfo {author} {\bibfnamefont {C.}~\bibnamefont {Cao}}, \bibinfo
  {author} {\bibfnamefont {X.}~\bibnamefont {Xu}}, \bibinfo {author}
  {\bibfnamefont {Z.}~\bibnamefont {Li}}, \bibinfo {author} {\bibfnamefont
  {D.}~\bibnamefont {Lv}},\ and\ \bibinfo {author} {\bibfnamefont {M.-H.}\
  \bibnamefont {Yung}},\ }\href@noop {} {\bibinfo {title} {Circuit-depth
  reduction of unitary-coupled-cluster ansatz by energy sorting}} (\bibinfo
  {year} {2021}),\ \Eprint {https://arxiv.org/abs/2106.15210}
  {arXiv:2106.15210} \BibitemShut {NoStop}%
\bibitem [{\citenamefont {Wecker}\ \emph {et~al.}(2015)\citenamefont {Wecker},
  \citenamefont {Hastings},\ and\ \citenamefont {Troyer}}]{wecker2015progress}%
  \BibitemOpen
  \bibfield  {author} {\bibinfo {author} {\bibfnamefont {D.}~\bibnamefont
  {Wecker}}, \bibinfo {author} {\bibfnamefont {M.~B.}\ \bibnamefont
  {Hastings}},\ and\ \bibinfo {author} {\bibfnamefont {M.}~\bibnamefont
  {Troyer}},\ }\href {https://doi.org/10.1103/PhysRevA.92.042303} {\bibfield
  {journal} {\bibinfo  {journal} {Phys. Rev. A}\ }\textbf {\bibinfo {volume}
  {92}},\ \bibinfo {pages} {042303} (\bibinfo {year} {2015})}\BibitemShut
  {NoStop}%
\bibitem [{\citenamefont {Ryabinkin}\ \emph {et~al.}(2018)\citenamefont
  {Ryabinkin}, \citenamefont {Yen}, \citenamefont {Genin},\ and\ \citenamefont
  {Izmaylov}}]{QCC}%
  \BibitemOpen
  \bibfield  {author} {\bibinfo {author} {\bibfnamefont {I.~G.}\ \bibnamefont
  {Ryabinkin}}, \bibinfo {author} {\bibfnamefont {T.-C.}\ \bibnamefont {Yen}},
  \bibinfo {author} {\bibfnamefont {S.~N.}\ \bibnamefont {Genin}},\ and\
  \bibinfo {author} {\bibfnamefont {A.~F.}\ \bibnamefont {Izmaylov}},\ }\href
  {https://doi.org/10.1021/acs.jctc.8b00932} {\bibfield  {journal} {\bibinfo
  {journal} {J. Chem. Theory Comput.}\ }\textbf {\bibinfo {volume} {14}},\
  \bibinfo {pages} {6317} (\bibinfo {year} {2018})}\BibitemShut {NoStop}%
\bibitem [{\citenamefont {Ryabinkin}\ \emph {et~al.}(2020)\citenamefont
  {Ryabinkin}, \citenamefont {Lang}, \citenamefont {Genin},\ and\ \citenamefont
  {Izmaylov}}]{Ryabinkin2020}%
  \BibitemOpen
  \bibfield  {author} {\bibinfo {author} {\bibfnamefont {I.~G.}\ \bibnamefont
  {Ryabinkin}}, \bibinfo {author} {\bibfnamefont {R.~A.}\ \bibnamefont {Lang}},
  \bibinfo {author} {\bibfnamefont {S.~N.}\ \bibnamefont {Genin}},\ and\
  \bibinfo {author} {\bibfnamefont {A.~F.}\ \bibnamefont {Izmaylov}},\ }\href
  {https://doi.org/10.1021/acs.jctc.9b01084} {\bibfield  {journal} {\bibinfo
  {journal} {J. Chem. Theory Comput.}\ }\textbf {\bibinfo {volume} {16}},\
  \bibinfo {pages} {1055} (\bibinfo {year} {2020})}\BibitemShut {NoStop}%
\bibitem [{\citenamefont {Dallaire-Demers}\ \emph {et~al.}(2019)\citenamefont
  {Dallaire-Demers}, \citenamefont {Romero}, \citenamefont {Veis},
  \citenamefont {Sim},\ and\ \citenamefont {Aspuru-Guzik}}]{dallaire2018low}%
  \BibitemOpen
  \bibfield  {author} {\bibinfo {author} {\bibfnamefont {P.~L.}\ \bibnamefont
  {Dallaire-Demers}}, \bibinfo {author} {\bibfnamefont {J.}~\bibnamefont
  {Romero}}, \bibinfo {author} {\bibfnamefont {L.}~\bibnamefont {Veis}},
  \bibinfo {author} {\bibfnamefont {S.}~\bibnamefont {Sim}},\ and\ \bibinfo
  {author} {\bibfnamefont {A.}~\bibnamefont {Aspuru-Guzik}},\ }\href
  {https://doi.org/10.1088/2058-9565/ab3951} {\bibfield  {journal} {\bibinfo
  {journal} {Quantum Sci. Technol.}\ }\textbf {\bibinfo {volume} {4}},\
  \bibinfo {pages} {1} (\bibinfo {year} {2019})}\BibitemShut {NoStop}%
\bibitem [{\citenamefont {Lee}\ \emph {et~al.}(2019)\citenamefont {Lee},
  \citenamefont {Huggins}, \citenamefont {Head-Gordon},\ and\ \citenamefont
  {Whaley}}]{Lee2019}%
  \BibitemOpen
  \bibfield  {author} {\bibinfo {author} {\bibfnamefont {J.}~\bibnamefont
  {Lee}}, \bibinfo {author} {\bibfnamefont {W.~J.}\ \bibnamefont {Huggins}},
  \bibinfo {author} {\bibfnamefont {M.}~\bibnamefont {Head-Gordon}},\ and\
  \bibinfo {author} {\bibfnamefont {K.~B.}\ \bibnamefont {Whaley}},\ }\href
  {https://doi.org/10.1021/acs.jctc.8b01004} {\bibfield  {journal} {\bibinfo
  {journal} {J. Chem. Theory Comput.}\ }\textbf {\bibinfo {volume} {15}},\
  \bibinfo {pages} {311} (\bibinfo {year} {2019})}\BibitemShut {NoStop}%
\bibitem [{\citenamefont {Bravyi}\ \emph {et~al.}(2017)\citenamefont {Bravyi},
  \citenamefont {Gambetta}, \citenamefont {Mezzacapo},\ and\ \citenamefont
  {Temme}}]{bravyi2017tapering}%
  \BibitemOpen
  \bibfield  {author} {\bibinfo {author} {\bibfnamefont {S.}~\bibnamefont
  {Bravyi}}, \bibinfo {author} {\bibfnamefont {J.~M.}\ \bibnamefont
  {Gambetta}}, \bibinfo {author} {\bibfnamefont {A.}~\bibnamefont
  {Mezzacapo}},\ and\ \bibinfo {author} {\bibfnamefont {K.}~\bibnamefont
  {Temme}},\ }\href@noop {} {\bibinfo {title} {Tapering off qubits to simulate
  fermionic hamiltonians}} (\bibinfo {year} {2017}),\ \Eprint
  {https://arxiv.org/abs/1701.08213} {arXiv:1701.08213} \BibitemShut {NoStop}%
\bibitem [{\citenamefont {Gard}\ \emph {et~al.}(2020)\citenamefont {Gard},
  \citenamefont {Zhu}, \citenamefont {Barron}, \citenamefont {Mayhall},
  \citenamefont {Economou},\ and\ \citenamefont {Barnes}}]{Gard2019}%
  \BibitemOpen
  \bibfield  {author} {\bibinfo {author} {\bibfnamefont {B.~T.}\ \bibnamefont
  {Gard}}, \bibinfo {author} {\bibfnamefont {L.}~\bibnamefont {Zhu}}, \bibinfo
  {author} {\bibfnamefont {G.~S.}\ \bibnamefont {Barron}}, \bibinfo {author}
  {\bibfnamefont {N.~J.}\ \bibnamefont {Mayhall}}, \bibinfo {author}
  {\bibfnamefont {S.~E.}\ \bibnamefont {Economou}},\ and\ \bibinfo {author}
  {\bibfnamefont {E.}~\bibnamefont {Barnes}},\ }\href
  {https://doi.org/10.1038/s41534-019-0240-1} {\bibfield  {journal} {\bibinfo
  {journal} {npj Quantum Inf.}\ }\textbf {\bibinfo {volume} {6}},\ \bibinfo
  {pages} {10} (\bibinfo {year} {2020})},\ \Eprint
  {https://arxiv.org/abs/1904.10910} {1904.10910} \BibitemShut {NoStop}%
\bibitem [{\citenamefont {Greene-Diniz}\ and\ \citenamefont
  {Muñoz~Ramo}(2021)}]{Greene-Diniz2021}%
  \BibitemOpen
  \bibfield  {author} {\bibinfo {author} {\bibfnamefont {G.}~\bibnamefont
  {Greene-Diniz}}\ and\ \bibinfo {author} {\bibfnamefont {D.}~\bibnamefont
  {Muñoz~Ramo}},\ }\href {https://doi.org/https://doi.org/10.1002/qua.26352}
  {\bibfield  {journal} {\bibinfo  {journal} {Int. J. Quantum Chem.}\ }\textbf
  {\bibinfo {volume} {121}},\ \bibinfo {pages} {e26352} (\bibinfo {year}
  {2021})}\BibitemShut {NoStop}%
\bibitem [{\citenamefont {Čársky}\ \emph {et~al.}(1987)\citenamefont
  {Čársky}, \citenamefont {Schaad}, \citenamefont {Hess}, \citenamefont
  {Urban},\ and\ \citenamefont {Noga}}]{CCsym_1}%
  \BibitemOpen
  \bibfield  {author} {\bibinfo {author} {\bibfnamefont {P.}~\bibnamefont
  {Čársky}}, \bibinfo {author} {\bibfnamefont {L.~J.}\ \bibnamefont
  {Schaad}}, \bibinfo {author} {\bibfnamefont {B.~A.}\ \bibnamefont {Hess}},
  \bibinfo {author} {\bibfnamefont {M.}~\bibnamefont {Urban}},\ and\ \bibinfo
  {author} {\bibfnamefont {J.}~\bibnamefont {Noga}},\ }\href
  {https://doi.org/10.1063/1.453585} {\bibfield  {journal} {\bibinfo  {journal}
  {J. Chem. Phys.}\ }\textbf {\bibinfo {volume} {87}},\ \bibinfo {pages} {411}
  (\bibinfo {year} {1987})}\BibitemShut {NoStop}%
\bibitem [{\citenamefont {Stanton}\ \emph {et~al.}(1991)\citenamefont
  {Stanton}, \citenamefont {Gauss}, \citenamefont {Watts},\ and\ \citenamefont
  {Bartlett}}]{CCsym_2}%
  \BibitemOpen
  \bibfield  {author} {\bibinfo {author} {\bibfnamefont {J.~F.}\ \bibnamefont
  {Stanton}}, \bibinfo {author} {\bibfnamefont {J.}~\bibnamefont {Gauss}},
  \bibinfo {author} {\bibfnamefont {J.~D.}\ \bibnamefont {Watts}},\ and\
  \bibinfo {author} {\bibfnamefont {R.~J.}\ \bibnamefont {Bartlett}},\ }\href
  {https://doi.org/10.1063/1.460620} {\bibfield  {journal} {\bibinfo  {journal}
  {J. Chem. Phys.}\ }\textbf {\bibinfo {volume} {94}},\ \bibinfo {pages} {4334}
  (\bibinfo {year} {1991})}\BibitemShut {NoStop}%
\bibitem [{\citenamefont {Scuseria}\ \emph {et~al.}(1987)\citenamefont
  {Scuseria}, \citenamefont {Scheiner}, \citenamefont {Lee}, \citenamefont
  {Rice},\ and\ \citenamefont {Schaefer}}]{CCSym3}%
  \BibitemOpen
  \bibfield  {author} {\bibinfo {author} {\bibfnamefont {G.~E.}\ \bibnamefont
  {Scuseria}}, \bibinfo {author} {\bibfnamefont {A.~C.}\ \bibnamefont
  {Scheiner}}, \bibinfo {author} {\bibfnamefont {T.~J.}\ \bibnamefont {Lee}},
  \bibinfo {author} {\bibfnamefont {J.~E.}\ \bibnamefont {Rice}},\ and\
  \bibinfo {author} {\bibfnamefont {H.~F.}\ \bibnamefont {Schaefer}},\ }\href
  {https://doi.org/10.1063/1.452039} {\bibfield  {journal} {\bibinfo  {journal}
  {J. Chem. Phys.}\ }\textbf {\bibinfo {volume} {86}},\ \bibinfo {pages} {2881}
  (\bibinfo {year} {1987})}\BibitemShut {NoStop}%
\bibitem [{\citenamefont {Setia}\ \emph {et~al.}(2020)\citenamefont {Setia},
  \citenamefont {Chen}, \citenamefont {Rice}, \citenamefont {Mezzacapo},
  \citenamefont {Pistoia},\ and\ \citenamefont
  {Whitfield}}]{setia2020reducing}%
  \BibitemOpen
  \bibfield  {author} {\bibinfo {author} {\bibfnamefont {K.}~\bibnamefont
  {Setia}}, \bibinfo {author} {\bibfnamefont {R.}~\bibnamefont {Chen}},
  \bibinfo {author} {\bibfnamefont {J.~E.}\ \bibnamefont {Rice}}, \bibinfo
  {author} {\bibfnamefont {A.}~\bibnamefont {Mezzacapo}}, \bibinfo {author}
  {\bibfnamefont {M.}~\bibnamefont {Pistoia}},\ and\ \bibinfo {author}
  {\bibfnamefont {J.~D.}\ \bibnamefont {Whitfield}},\ }\href
  {https://doi.org/10.1021/acs.jctc.0c00113} {\bibfield  {journal} {\bibinfo
  {journal} {J. Chem. Theory Comput.}\ }\textbf {\bibinfo {volume} {16}},\
  \bibinfo {pages} {6091} (\bibinfo {year} {2020})}\BibitemShut {NoStop}%
\bibitem [{\citenamefont {Fischer}\ and\ \citenamefont
  {Gunlycke}(2019)}]{fischer2019symmetry}%
  \BibitemOpen
  \bibfield  {author} {\bibinfo {author} {\bibfnamefont {S.~A.}\ \bibnamefont
  {Fischer}}\ and\ \bibinfo {author} {\bibfnamefont {D.}~\bibnamefont
  {Gunlycke}},\ }\href@noop {} {\bibinfo {title} {Symmetry configuration
  mapping for representing quantum systems on quantum computers}} (\bibinfo
  {year} {2019}),\ \Eprint {https://arxiv.org/abs/1907.01493}
  {arXiv:1907.01493} \BibitemShut {NoStop}%
\bibitem [{\citenamefont {Seki}\ \emph {et~al.}(2020)\citenamefont {Seki},
  \citenamefont {Shirakawa},\ and\ \citenamefont
  {Yunoki}}]{PhysRevA.101.052340}%
  \BibitemOpen
  \bibfield  {author} {\bibinfo {author} {\bibfnamefont {K.}~\bibnamefont
  {Seki}}, \bibinfo {author} {\bibfnamefont {T.}~\bibnamefont {Shirakawa}},\
  and\ \bibinfo {author} {\bibfnamefont {S.}~\bibnamefont {Yunoki}},\ }\href
  {https://doi.org/10.1103/PhysRevA.101.052340} {\bibfield  {journal} {\bibinfo
   {journal} {Phys. Rev. A}\ }\textbf {\bibinfo {volume} {101}},\ \bibinfo
  {pages} {052340} (\bibinfo {year} {2020})}\BibitemShut {NoStop}%
\bibitem [{\citenamefont {Yen}\ \emph {et~al.}(2019)\citenamefont {Yen},
  \citenamefont {Lang},\ and\ \citenamefont {Izmaylov}}]{hamiton_sym2}%
  \BibitemOpen
  \bibfield  {author} {\bibinfo {author} {\bibfnamefont {T.-C.}\ \bibnamefont
  {Yen}}, \bibinfo {author} {\bibfnamefont {R.~A.}\ \bibnamefont {Lang}},\ and\
  \bibinfo {author} {\bibfnamefont {A.~F.}\ \bibnamefont {Izmaylov}},\ }\href
  {https://doi.org/10.1063/1.5110682} {\bibfield  {journal} {\bibinfo
  {journal} {J. Chem. Phys.}\ }\textbf {\bibinfo {volume} {151}},\ \bibinfo
  {pages} {164111} (\bibinfo {year} {2019})}\BibitemShut {NoStop}%
\bibitem [{\citenamefont {McArdle}\ \emph
  {et~al.}(2020{\natexlab{b}})\citenamefont {McArdle}, \citenamefont {Endo},
  \citenamefont {Aspuru-Guzik}, \citenamefont {Benjamin},\ and\ \citenamefont
  {Yuan}}]{McArdle2018}%
  \BibitemOpen
  \bibfield  {author} {\bibinfo {author} {\bibfnamefont {S.}~\bibnamefont
  {McArdle}}, \bibinfo {author} {\bibfnamefont {S.}~\bibnamefont {Endo}},
  \bibinfo {author} {\bibfnamefont {A.}~\bibnamefont {Aspuru-Guzik}}, \bibinfo
  {author} {\bibfnamefont {S.~C.}\ \bibnamefont {Benjamin}},\ and\ \bibinfo
  {author} {\bibfnamefont {X.}~\bibnamefont {Yuan}},\ }\href
  {https://doi.org/10.1103/RevModPhys.92.015003} {\bibfield  {journal}
  {\bibinfo  {journal} {Rev. Mod. Phys.}\ }\textbf {\bibinfo {volume} {92}},\
  \bibinfo {pages} {015003} (\bibinfo {year} {2020}{\natexlab{b}})}\BibitemShut
  {NoStop}%
\bibitem [{\citenamefont {Hoffmann}\ and\ \citenamefont
  {Simons}(1988)}]{hoffmann1988unitary}%
  \BibitemOpen
  \bibfield  {author} {\bibinfo {author} {\bibfnamefont {M.~R.}\ \bibnamefont
  {Hoffmann}}\ and\ \bibinfo {author} {\bibfnamefont {J.}~\bibnamefont
  {Simons}},\ }\href {https://doi.org/10.1063/1.454125} {\bibfield  {journal}
  {\bibinfo  {journal} {J. Chem. Phys.}\ }\textbf {\bibinfo {volume} {88}},\
  \bibinfo {pages} {993} (\bibinfo {year} {1988})}\BibitemShut {NoStop}%
\bibitem [{\citenamefont {Bartlett}\ \emph {et~al.}(1989)\citenamefont
  {Bartlett}, \citenamefont {Kucharski},\ and\ \citenamefont
  {Noga}}]{bartlett1989alternative}%
  \BibitemOpen
  \bibfield  {author} {\bibinfo {author} {\bibfnamefont {R.~J.}\ \bibnamefont
  {Bartlett}}, \bibinfo {author} {\bibfnamefont {S.~A.}\ \bibnamefont
  {Kucharski}},\ and\ \bibinfo {author} {\bibfnamefont {J.}~\bibnamefont
  {Noga}},\ }\href
  {https://doi.org/https://doi.org/10.1016/S0009-2614(89)87372-5} {\bibfield
  {journal} {\bibinfo  {journal} {Chem. Phys. Lett.}\ }\textbf {\bibinfo
  {volume} {155}},\ \bibinfo {pages} {133} (\bibinfo {year}
  {1989})}\BibitemShut {NoStop}%
\bibitem [{\citenamefont {Shen}\ \emph {et~al.}(2017)\citenamefont {Shen},
  \citenamefont {Zhang}, \citenamefont {Zhang}, \citenamefont {Zhang},
  \citenamefont {Yung},\ and\ \citenamefont {Kim}}]{shen2017quantum}%
  \BibitemOpen
  \bibfield  {author} {\bibinfo {author} {\bibfnamefont {Y.}~\bibnamefont
  {Shen}}, \bibinfo {author} {\bibfnamefont {X.}~\bibnamefont {Zhang}},
  \bibinfo {author} {\bibfnamefont {S.}~\bibnamefont {Zhang}}, \bibinfo
  {author} {\bibfnamefont {J.-N.}\ \bibnamefont {Zhang}}, \bibinfo {author}
  {\bibfnamefont {M.-H.}\ \bibnamefont {Yung}},\ and\ \bibinfo {author}
  {\bibfnamefont {K.}~\bibnamefont {Kim}},\ }\href
  {https://doi.org/10.1103/PhysRevA.95.020501} {\bibfield  {journal} {\bibinfo
  {journal} {Phys. Rev. A}\ }\textbf {\bibinfo {volume} {95}},\ \bibinfo
  {pages} {020501} (\bibinfo {year} {2017})}\BibitemShut {NoStop}%
\bibitem [{\citenamefont {Liu}\ \emph {et~al.}(2021)\citenamefont {Liu},
  \citenamefont {Sun},\ and\ \citenamefont {Yuan}}]{liu2021variational}%
  \BibitemOpen
  \bibfield  {author} {\bibinfo {author} {\bibfnamefont {J.}~\bibnamefont
  {Liu}}, \bibinfo {author} {\bibfnamefont {J.}~\bibnamefont {Sun}},\ and\
  \bibinfo {author} {\bibfnamefont {X.}~\bibnamefont {Yuan}},\ }\href@noop {}
  {\bibinfo {title} {Towards a variational jordan-lee-preskill quantum
  algorithm}} (\bibinfo {year} {2021}),\ \Eprint
  {https://arxiv.org/abs/2109.05547} {arXiv:2109.05547} \BibitemShut {NoStop}%
\bibitem [{\citenamefont {Barkoutsos}\ \emph {et~al.}(2018)\citenamefont
  {Barkoutsos}, \citenamefont {Gonthier}, \citenamefont {Sokolov},
  \citenamefont {Moll}, \citenamefont {Salis}, \citenamefont {Fuhrer},
  \citenamefont {Ganzhorn}, \citenamefont {Egger}, \citenamefont {Troyer},
  \citenamefont {Mezzacapo}, \citenamefont {Filipp},\ and\ \citenamefont
  {Tavernelli}}]{PRA_trott_discuss}%
  \BibitemOpen
  \bibfield  {author} {\bibinfo {author} {\bibfnamefont {P.~K.}\ \bibnamefont
  {Barkoutsos}}, \bibinfo {author} {\bibfnamefont {J.~F.}\ \bibnamefont
  {Gonthier}}, \bibinfo {author} {\bibfnamefont {I.}~\bibnamefont {Sokolov}},
  \bibinfo {author} {\bibfnamefont {N.}~\bibnamefont {Moll}}, \bibinfo {author}
  {\bibfnamefont {G.}~\bibnamefont {Salis}}, \bibinfo {author} {\bibfnamefont
  {A.}~\bibnamefont {Fuhrer}}, \bibinfo {author} {\bibfnamefont
  {M.}~\bibnamefont {Ganzhorn}}, \bibinfo {author} {\bibfnamefont {D.~J.}\
  \bibnamefont {Egger}}, \bibinfo {author} {\bibfnamefont {M.}~\bibnamefont
  {Troyer}}, \bibinfo {author} {\bibfnamefont {A.}~\bibnamefont {Mezzacapo}},
  \bibinfo {author} {\bibfnamefont {S.}~\bibnamefont {Filipp}},\ and\ \bibinfo
  {author} {\bibfnamefont {I.}~\bibnamefont {Tavernelli}},\ }\href
  {https://doi.org/10.1103/PhysRevA.98.022322} {\bibfield  {journal} {\bibinfo
  {journal} {Phys. Rev. A}\ }\textbf {\bibinfo {volume} {98}},\ \bibinfo
  {pages} {022322} (\bibinfo {year} {2018})}\BibitemShut {NoStop}%
\bibitem [{\citenamefont {Sun}\ \emph {et~al.}(2020)\citenamefont {Sun},
  \citenamefont {Zhang}, \citenamefont {Banerjee}, \citenamefont {Bao},
  \citenamefont {Barbry}, \citenamefont {Blunt}, \citenamefont {Bogdanov},
  \citenamefont {Wang}, \citenamefont {White},\ and\ \citenamefont
  {Whitfield}}]{PySCF_1}%
  \BibitemOpen
  \bibfield  {author} {\bibinfo {author} {\bibfnamefont {Q.}~\bibnamefont
  {Sun}}, \bibinfo {author} {\bibfnamefont {X.}~\bibnamefont {Zhang}}, \bibinfo
  {author} {\bibfnamefont {S.}~\bibnamefont {Banerjee}}, \bibinfo {author}
  {\bibfnamefont {P.}~\bibnamefont {Bao}}, \bibinfo {author} {\bibfnamefont
  {M.}~\bibnamefont {Barbry}}, \bibinfo {author} {\bibfnamefont {N.~S.}\
  \bibnamefont {Blunt}}, \bibinfo {author} {\bibfnamefont {N.~A.}\ \bibnamefont
  {Bogdanov}}, \bibinfo {author} {\bibfnamefont {X.}~\bibnamefont {Wang}},
  \bibinfo {author} {\bibfnamefont {A.}~\bibnamefont {White}},\ and\ \bibinfo
  {author} {\bibfnamefont {J.~D. e.~a.}\ \bibnamefont {Whitfield}},\ }\href
  {https://doi.org/10.1063/5.0006074} {\bibfield  {journal} {\bibinfo
  {journal} {J. Chem. Phys.}\ }\textbf {\bibinfo {volume} {153}},\ \bibinfo
  {pages} {024109} (\bibinfo {year} {2020})}\BibitemShut {NoStop}%
\bibitem [{\citenamefont {Sun}\ \emph {et~al.}(2018)\citenamefont {Sun},
  \citenamefont {Berkelbach}, \citenamefont {Blunt}, \citenamefont {Booth},
  \citenamefont {Guo}, \citenamefont {Li}, \citenamefont {Liu}, \citenamefont
  {McClain}, \citenamefont {Sayfutyarova},\ and\ \citenamefont
  {Sharma}}]{PySCF_2}%
  \BibitemOpen
  \bibfield  {author} {\bibinfo {author} {\bibfnamefont {Q.}~\bibnamefont
  {Sun}}, \bibinfo {author} {\bibfnamefont {T.~C.}\ \bibnamefont {Berkelbach}},
  \bibinfo {author} {\bibfnamefont {N.~S.}\ \bibnamefont {Blunt}}, \bibinfo
  {author} {\bibfnamefont {G.~H.}\ \bibnamefont {Booth}}, \bibinfo {author}
  {\bibfnamefont {S.}~\bibnamefont {Guo}}, \bibinfo {author} {\bibfnamefont
  {Z.}~\bibnamefont {Li}}, \bibinfo {author} {\bibfnamefont {J.}~\bibnamefont
  {Liu}}, \bibinfo {author} {\bibfnamefont {J.~D.}\ \bibnamefont {McClain}},
  \bibinfo {author} {\bibfnamefont {E.~R.}\ \bibnamefont {Sayfutyarova}},\ and\
  \bibinfo {author} {\bibfnamefont {S.~e.~a.}\ \bibnamefont {Sharma}},\ }\href
  {https://doi.org/https://doi.org/10.1002/wcms.1340} {\bibfield  {journal}
  {\bibinfo  {journal} {Wiley Interdiscip. Rev. Comput. Mol. Sci.}\ }\textbf
  {\bibinfo {volume} {8}},\ \bibinfo {pages} {e1340} (\bibinfo {year}
  {2018})}\BibitemShut {NoStop}%
\bibitem [{\citenamefont {Huawei}(2021)}]{mindquantum}%
  \BibitemOpen
  \bibfield  {author} {\bibinfo {author} {\bibnamefont {Huawei}},\ }\href@noop
  {} {\bibinfo {title} {Mindquantum}},\ \bibinfo {howpublished} {Website}
  (\bibinfo {year} {2021}),\ \bibinfo {note}
  {\url{https://gitee.com/mindspore/mindquantum}}\BibitemShut {NoStop}%
\bibitem [{\citenamefont {Hehre}\ \emph {et~al.}(1970)\citenamefont {Hehre},
  \citenamefont {Ditchfield}, \citenamefont {Stewart},\ and\ \citenamefont
  {Pople}}]{sto3g}%
  \BibitemOpen
  \bibfield  {author} {\bibinfo {author} {\bibfnamefont {W.~J.}\ \bibnamefont
  {Hehre}}, \bibinfo {author} {\bibfnamefont {R.}~\bibnamefont {Ditchfield}},
  \bibinfo {author} {\bibfnamefont {R.~F.}\ \bibnamefont {Stewart}},\ and\
  \bibinfo {author} {\bibfnamefont {J.~A.}\ \bibnamefont {Pople}},\ }\href
  {https://doi.org/10.1063/1.1673374} {\bibfield  {journal} {\bibinfo
  {journal} {The Journal of Chemical Physics}\ }\textbf {\bibinfo {volume}
  {52}},\ \bibinfo {pages} {2769} (\bibinfo {year} {1970})}\BibitemShut
  {NoStop}%
\bibitem [{\citenamefont {Jordan}\ and\ \citenamefont
  {Wigner}(1993)}]{Jordan1993}%
  \BibitemOpen
  \bibfield  {author} {\bibinfo {author} {\bibfnamefont {P.}~\bibnamefont
  {Jordan}}\ and\ \bibinfo {author} {\bibfnamefont {E.~P.}\ \bibnamefont
  {Wigner}},\ }\href {https://doi.org/10.1007/978-3-662-02781-3_9} {\bibfield
  {journal} {\bibinfo  {journal} {The Collected Works of Eugene Paul Wigner}\
  ,\ \bibinfo {pages} {109}} (\bibinfo {year} {1993})}\BibitemShut {NoStop}%
\bibitem [{\citenamefont {Russell~Johnson}(2020)}]{NISTCCCBDB}%
  \BibitemOpen
  \bibfield  {author} {\bibinfo {author} {\bibfnamefont {N.}~\bibnamefont
  {Russell~Johnson}},\ }\href {https://cccbdb.nist.gov/} {\bibinfo {title}
  {{NIST} computational chemistry comparison and benchmark database}} (\bibinfo
  {year} {2020})\BibitemShut {NoStop}%
\bibitem [{\citenamefont {Grimsley}\ \emph
  {et~al.}(2019{\natexlab{b}})\citenamefont {Grimsley}, \citenamefont
  {Economou}, \citenamefont {Barnes},\ and\ \citenamefont
  {Mayhall}}]{adaptvqe}%
  \BibitemOpen
  \bibfield  {author} {\bibinfo {author} {\bibfnamefont {H.~R.}\ \bibnamefont
  {Grimsley}}, \bibinfo {author} {\bibfnamefont {S.~E.}\ \bibnamefont
  {Economou}}, \bibinfo {author} {\bibfnamefont {E.}~\bibnamefont {Barnes}},\
  and\ \bibinfo {author} {\bibfnamefont {N.~J.}\ \bibnamefont {Mayhall}},\
  }\href {https://doi.org/10.1038/s41467-019-10988-2} {\bibfield  {journal}
  {\bibinfo  {journal} {Nat. Commun.}\ }\textbf {\bibinfo {volume} {10}},\
  \bibinfo {pages} {1} (\bibinfo {year} {2019}{\natexlab{b}})}\BibitemShut
  {NoStop}%
\bibitem [{\citenamefont {Perdew}\ \emph {et~al.}(1996)\citenamefont {Perdew},
  \citenamefont {Burke},\ and\ \citenamefont {Ernzerhof}}]{PBE_DFT}%
  \BibitemOpen
  \bibfield  {author} {\bibinfo {author} {\bibfnamefont {J.~P.}\ \bibnamefont
  {Perdew}}, \bibinfo {author} {\bibfnamefont {K.}~\bibnamefont {Burke}},\ and\
  \bibinfo {author} {\bibfnamefont {M.}~\bibnamefont {Ernzerhof}},\ }\href
  {https://doi.org/10.1103/PhysRevLett.77.3865} {\bibfield  {journal} {\bibinfo
   {journal} {Phys. Rev. Lett.}\ }\textbf {\bibinfo {volume} {77}},\ \bibinfo
  {pages} {3865} (\bibinfo {year} {1996})}\BibitemShut {NoStop}%
\bibitem [{\citenamefont {Weigend}\ and\ \citenamefont
  {Ahlrichs}(2005)}]{def2tzvp_basisset}%
  \BibitemOpen
  \bibfield  {author} {\bibinfo {author} {\bibfnamefont {F.}~\bibnamefont
  {Weigend}}\ and\ \bibinfo {author} {\bibfnamefont {R.}~\bibnamefont
  {Ahlrichs}},\ }\href {https://doi.org/10.1039/B508541A} {\bibfield  {journal}
  {\bibinfo  {journal} {Phys. Chem. Chem. Phys.}\ }\textbf {\bibinfo {volume}
  {7}},\ \bibinfo {pages} {3297} (\bibinfo {year} {2005})}\BibitemShut
  {NoStop}%
\bibitem [{\citenamefont {Holmes}\ \emph {et~al.}(2022)\citenamefont {Holmes},
  \citenamefont {Sharma}, \citenamefont {Cerezo},\ and\ \citenamefont
  {Coles}}]{holmes2022connecting}%
  \BibitemOpen
  \bibfield  {author} {\bibinfo {author} {\bibfnamefont {Z.}~\bibnamefont
  {Holmes}}, \bibinfo {author} {\bibfnamefont {K.}~\bibnamefont {Sharma}},
  \bibinfo {author} {\bibfnamefont {M.}~\bibnamefont {Cerezo}},\ and\ \bibinfo
  {author} {\bibfnamefont {P.~J.}\ \bibnamefont {Coles}},\ }\href@noop {}
  {\bibfield  {journal} {\bibinfo  {journal} {PRX Quantum}\ }\textbf {\bibinfo
  {volume} {3}},\ \bibinfo {pages} {010313} (\bibinfo {year}
  {2022})}\BibitemShut {NoStop}%
\bibitem [{\citenamefont {Larocca}\ \emph {et~al.}(2021)\citenamefont
  {Larocca}, \citenamefont {Ju}, \citenamefont {Garc{\'\i}a-Mart{\'\i}n},
  \citenamefont {Coles},\ and\ \citenamefont {Cerezo}}]{larocca2021theory}%
  \BibitemOpen
  \bibfield  {author} {\bibinfo {author} {\bibfnamefont {M.}~\bibnamefont
  {Larocca}}, \bibinfo {author} {\bibfnamefont {N.}~\bibnamefont {Ju}},
  \bibinfo {author} {\bibfnamefont {D.}~\bibnamefont
  {Garc{\'\i}a-Mart{\'\i}n}}, \bibinfo {author} {\bibfnamefont {P.~J.}\
  \bibnamefont {Coles}},\ and\ \bibinfo {author} {\bibfnamefont
  {M.}~\bibnamefont {Cerezo}},\ }\href@noop {} {\bibfield  {journal} {\bibinfo
  {journal} {arXiv preprint arXiv:2109.11676}\ } (\bibinfo {year}
  {2021})}\BibitemShut {NoStop}%
\bibitem [{\citenamefont {Motta}\ \emph {et~al.}(2021)\citenamefont {Motta},
  \citenamefont {Ye}, \citenamefont {McClean}, \citenamefont {Li},
  \citenamefont {Minnich}, \citenamefont {Babbush},\ and\ \citenamefont
  {Chan}}]{motta2021low}%
  \BibitemOpen
  \bibfield  {author} {\bibinfo {author} {\bibfnamefont {M.}~\bibnamefont
  {Motta}}, \bibinfo {author} {\bibfnamefont {E.}~\bibnamefont {Ye}}, \bibinfo
  {author} {\bibfnamefont {J.~R.}\ \bibnamefont {McClean}}, \bibinfo {author}
  {\bibfnamefont {Z.}~\bibnamefont {Li}}, \bibinfo {author} {\bibfnamefont
  {A.~J.}\ \bibnamefont {Minnich}}, \bibinfo {author} {\bibfnamefont
  {R.}~\bibnamefont {Babbush}},\ and\ \bibinfo {author} {\bibfnamefont
  {G.~K.-L.}\ \bibnamefont {Chan}},\ }\href
  {https://doi.org/https://doi.org/10.1038/s41534-021-00416-z} {\bibfield
  {journal} {\bibinfo  {journal} {npj Quantum Information}\ }\textbf {\bibinfo
  {volume} {7}},\ \bibinfo {pages} {1} (\bibinfo {year} {2021})}\BibitemShut
  {NoStop}%
\bibitem [{\citenamefont {Matsuzawa}\ and\ \citenamefont
  {Kurashige}(2020)}]{jastrow2020}%
  \BibitemOpen
  \bibfield  {author} {\bibinfo {author} {\bibfnamefont {Y.}~\bibnamefont
  {Matsuzawa}}\ and\ \bibinfo {author} {\bibfnamefont {Y.}~\bibnamefont
  {Kurashige}},\ }\href {https://doi.org/https://10.1021/acs.jctc.9b00963}
  {\bibfield  {journal} {\bibinfo  {journal} {J. Chem. Theory Comput.}\
  }\textbf {\bibinfo {volume} {16}},\ \bibinfo {pages} {944} (\bibinfo {year}
  {2020})}\BibitemShut {NoStop}%
\bibitem [{\citenamefont {Rubin}\ \emph {et~al.}(2021)\citenamefont {Rubin},
  \citenamefont {Lee},\ and\ \citenamefont {Babbush}}]{rubin2021compressing}%
  \BibitemOpen
  \bibfield  {author} {\bibinfo {author} {\bibfnamefont {N.~C.}\ \bibnamefont
  {Rubin}}, \bibinfo {author} {\bibfnamefont {J.}~\bibnamefont {Lee}},\ and\
  \bibinfo {author} {\bibfnamefont {R.}~\bibnamefont {Babbush}},\ }\href@noop
  {} {\bibinfo {title} {Compressing many-body fermion operators under unitary
  constraints}} (\bibinfo {year} {2021}),\ \Eprint
  {https://arxiv.org/abs/2109.05010} {arXiv:2109.05010} \BibitemShut {NoStop}%
\bibitem [{\citenamefont {Babbush}\ \emph
  {et~al.}(2018{\natexlab{b}})\citenamefont {Babbush}, \citenamefont {Wiebe},
  \citenamefont {McClean}, \citenamefont {McClain}, \citenamefont {Neven},\
  and\ \citenamefont {Chan}}]{PhysRevXlow_depth}%
  \BibitemOpen
  \bibfield  {author} {\bibinfo {author} {\bibfnamefont {R.}~\bibnamefont
  {Babbush}}, \bibinfo {author} {\bibfnamefont {N.}~\bibnamefont {Wiebe}},
  \bibinfo {author} {\bibfnamefont {J.}~\bibnamefont {McClean}}, \bibinfo
  {author} {\bibfnamefont {J.}~\bibnamefont {McClain}}, \bibinfo {author}
  {\bibfnamefont {H.}~\bibnamefont {Neven}},\ and\ \bibinfo {author}
  {\bibfnamefont {G.~K.-L.}\ \bibnamefont {Chan}},\ }\href
  {https://doi.org/10.1103/PhysRevX.8.011044} {\bibfield  {journal} {\bibinfo
  {journal} {Phys. Rev. X}\ }\textbf {\bibinfo {volume} {8}},\ \bibinfo {pages}
  {011044} (\bibinfo {year} {2018}{\natexlab{b}})}\BibitemShut {NoStop}%
\bibitem [{\citenamefont {Kottmann}\ and\ \citenamefont
  {Aspuru-Guzik}(2021)}]{kottmann2021optimize}%
  \BibitemOpen
  \bibfield  {author} {\bibinfo {author} {\bibfnamefont {J.~S.}\ \bibnamefont
  {Kottmann}}\ and\ \bibinfo {author} {\bibfnamefont {A.}~\bibnamefont
  {Aspuru-Guzik}},\ }\href@noop {} {\bibinfo {title} {Optimized low-depth
  quantum circuits for molecular electronic structure using a separable pair
  approximation}} (\bibinfo {year} {2021}),\ \Eprint
  {https://arxiv.org/abs/2105.03836} {arXiv:2105.03836} \BibitemShut {NoStop}%
\bibitem [{\citenamefont {Tang}\ \emph {et~al.}(2021)\citenamefont {Tang},
  \citenamefont {Shkolnikov}, \citenamefont {Barron}, \citenamefont {Grimsley},
  \citenamefont {Mayhall}, \citenamefont {Barnes},\ and\ \citenamefont
  {Economou}}]{tang2021qubit}%
  \BibitemOpen
  \bibfield  {author} {\bibinfo {author} {\bibfnamefont {H.~L.}\ \bibnamefont
  {Tang}}, \bibinfo {author} {\bibfnamefont {V.}~\bibnamefont {Shkolnikov}},
  \bibinfo {author} {\bibfnamefont {G.~S.}\ \bibnamefont {Barron}}, \bibinfo
  {author} {\bibfnamefont {H.~R.}\ \bibnamefont {Grimsley}}, \bibinfo {author}
  {\bibfnamefont {N.~J.}\ \bibnamefont {Mayhall}}, \bibinfo {author}
  {\bibfnamefont {E.}~\bibnamefont {Barnes}},\ and\ \bibinfo {author}
  {\bibfnamefont {S.~E.}\ \bibnamefont {Economou}},\ }\href
  {https://doi.org/10.1103/PRXQuantum.2.020310} {\bibfield  {journal} {\bibinfo
   {journal} {PRX Quantum}\ }\textbf {\bibinfo {volume} {2}},\ \bibinfo {pages}
  {020310} (\bibinfo {year} {2021})}\BibitemShut {NoStop}%
\bibitem [{\citenamefont {Zhang}\ \emph {et~al.}(2020)\citenamefont {Zhang},
  \citenamefont {Sun}, \citenamefont {Yuan},\ and\ \citenamefont
  {Yung}}]{zhang2020lowdepth}%
  \BibitemOpen
  \bibfield  {author} {\bibinfo {author} {\bibfnamefont {Z.-J.}\ \bibnamefont
  {Zhang}}, \bibinfo {author} {\bibfnamefont {J.}~\bibnamefont {Sun}}, \bibinfo
  {author} {\bibfnamefont {X.}~\bibnamefont {Yuan}},\ and\ \bibinfo {author}
  {\bibfnamefont {M.-H.}\ \bibnamefont {Yung}},\ }\href@noop {} {\bibinfo
  {title} {Low-depth hamiltonian simulation by adaptive product formula}}
  (\bibinfo {year} {2020}),\ \Eprint {https://arxiv.org/abs/2011.05283}
  {arXiv:2011.05283} \BibitemShut {NoStop}%
\bibitem [{\citenamefont {Trout}\ and\ \citenamefont
  {Brown}(2015)}]{Trout_2015}%
  \BibitemOpen
  \bibfield  {author} {\bibinfo {author} {\bibfnamefont {C.~J.}\ \bibnamefont
  {Trout}}\ and\ \bibinfo {author} {\bibfnamefont {K.~R.}\ \bibnamefont
  {Brown}},\ }\href {https://doi.org/10.1002/qua.24856} {\bibfield  {journal}
  {\bibinfo  {journal} {Int. J. Quantum Chem.}\ }\textbf {\bibinfo {volume}
  {115}},\ \bibinfo {pages} {1296–1304} (\bibinfo {year} {2015})}\BibitemShut
  {NoStop}%
\bibitem [{\citenamefont {Jones}\ and\ \citenamefont
  {Benjamin}(2020)}]{jones2020quantum}%
  \BibitemOpen
  \bibfield  {author} {\bibinfo {author} {\bibfnamefont {T.}~\bibnamefont
  {Jones}}\ and\ \bibinfo {author} {\bibfnamefont {S.~C.}\ \bibnamefont
  {Benjamin}},\ }\href@noop {} {\bibinfo {title} {Quantum compilation and
  circuit optimisation via energy dissipation}} (\bibinfo {year} {2020}),\
  \Eprint {https://arxiv.org/abs/1811.03147} {arXiv:1811.03147} \BibitemShut
  {NoStop}%
\bibitem [{\citenamefont {Sun}\ \emph {et~al.}(2021)\citenamefont {Sun},
  \citenamefont {Endo}, \citenamefont {Lin}, \citenamefont {Hayden},
  \citenamefont {Vedral},\ and\ \citenamefont {Yuan}}]{sun2021perturbative}%
  \BibitemOpen
  \bibfield  {author} {\bibinfo {author} {\bibfnamefont {J.}~\bibnamefont
  {Sun}}, \bibinfo {author} {\bibfnamefont {S.}~\bibnamefont {Endo}}, \bibinfo
  {author} {\bibfnamefont {H.}~\bibnamefont {Lin}}, \bibinfo {author}
  {\bibfnamefont {P.}~\bibnamefont {Hayden}}, \bibinfo {author} {\bibfnamefont
  {V.}~\bibnamefont {Vedral}},\ and\ \bibinfo {author} {\bibfnamefont
  {X.}~\bibnamefont {Yuan}},\ }\href@noop {} {\bibinfo {title} {Perturbative
  quantum simulation}} (\bibinfo {year} {2021}),\ \Eprint
  {https://arxiv.org/abs/2106.05938} {arXiv:2106.05938} \BibitemShut {NoStop}%
\bibitem [{\citenamefont {Fujii}\ \emph {et~al.}(2020)\citenamefont {Fujii},
  \citenamefont {Mitarai}, \citenamefont {Mizukami},\ and\ \citenamefont
  {Nakagawa}}]{fujii2020deep}%
  \BibitemOpen
  \bibfield  {author} {\bibinfo {author} {\bibfnamefont {K.}~\bibnamefont
  {Fujii}}, \bibinfo {author} {\bibfnamefont {K.}~\bibnamefont {Mitarai}},
  \bibinfo {author} {\bibfnamefont {W.}~\bibnamefont {Mizukami}},\ and\
  \bibinfo {author} {\bibfnamefont {Y.~O.}\ \bibnamefont {Nakagawa}},\
  }\href@noop {} {\bibinfo {title} {Deep variational quantum eigensolver: a
  divide-and-conquer method for solving a larger problem with smaller size
  quantum computers}} (\bibinfo {year} {2020}),\ \Eprint
  {https://arxiv.org/abs/2007.10917} {arXiv:2007.10917} \BibitemShut {NoStop}%
\bibitem [{\citenamefont {Mizuta}\ \emph {et~al.}(2021)\citenamefont {Mizuta},
  \citenamefont {Fujii}, \citenamefont {Fujii}, \citenamefont {Ichikawa},
  \citenamefont {Imamura}, \citenamefont {Okuno},\ and\ \citenamefont
  {Nakagawa}}]{mizuta2021deep}%
  \BibitemOpen
  \bibfield  {author} {\bibinfo {author} {\bibfnamefont {K.}~\bibnamefont
  {Mizuta}}, \bibinfo {author} {\bibfnamefont {M.}~\bibnamefont {Fujii}},
  \bibinfo {author} {\bibfnamefont {S.}~\bibnamefont {Fujii}}, \bibinfo
  {author} {\bibfnamefont {K.}~\bibnamefont {Ichikawa}}, \bibinfo {author}
  {\bibfnamefont {Y.}~\bibnamefont {Imamura}}, \bibinfo {author} {\bibfnamefont
  {Y.}~\bibnamefont {Okuno}},\ and\ \bibinfo {author} {\bibfnamefont {Y.~O.}\
  \bibnamefont {Nakagawa}},\ }\href@noop {} {\bibinfo {title} {Deep variational
  quantum eigensolver for excited states and its application to quantum
  chemistry calculation of periodic materials}} (\bibinfo {year} {2021}),\
  \Eprint {https://arxiv.org/abs/2104.00855} {arXiv:2104.00855} \BibitemShut
  {NoStop}%
\bibitem [{\citenamefont {Takeshita}\ \emph {et~al.}(2020)\citenamefont
  {Takeshita}, \citenamefont {Rubin}, \citenamefont {Jiang}, \citenamefont
  {Lee}, \citenamefont {Babbush},\ and\ \citenamefont
  {McClean}}]{takeshita2020increasing}%
  \BibitemOpen
  \bibfield  {author} {\bibinfo {author} {\bibfnamefont {T.}~\bibnamefont
  {Takeshita}}, \bibinfo {author} {\bibfnamefont {N.~C.}\ \bibnamefont
  {Rubin}}, \bibinfo {author} {\bibfnamefont {Z.}~\bibnamefont {Jiang}},
  \bibinfo {author} {\bibfnamefont {E.}~\bibnamefont {Lee}}, \bibinfo {author}
  {\bibfnamefont {R.}~\bibnamefont {Babbush}},\ and\ \bibinfo {author}
  {\bibfnamefont {J.~R.}\ \bibnamefont {McClean}},\ }\href
  {https://doi.org/10.1103/PhysRevX.10.011004} {\bibfield  {journal} {\bibinfo
  {journal} {Phys. Rev. X}\ }\textbf {\bibinfo {volume} {10}},\ \bibinfo
  {pages} {011004} (\bibinfo {year} {2020})}\BibitemShut {NoStop}%
\bibitem [{\citenamefont {Yuan}\ \emph {et~al.}(2021)\citenamefont {Yuan},
  \citenamefont {Sun}, \citenamefont {Liu}, \citenamefont {Zhao},\ and\
  \citenamefont {Zhou}}]{yuan2021quantum}%
  \BibitemOpen
  \bibfield  {author} {\bibinfo {author} {\bibfnamefont {X.}~\bibnamefont
  {Yuan}}, \bibinfo {author} {\bibfnamefont {J.}~\bibnamefont {Sun}}, \bibinfo
  {author} {\bibfnamefont {J.}~\bibnamefont {Liu}}, \bibinfo {author}
  {\bibfnamefont {Q.}~\bibnamefont {Zhao}},\ and\ \bibinfo {author}
  {\bibfnamefont {Y.}~\bibnamefont {Zhou}},\ }\href
  {https://doi.org/10.1103/PhysRevLett.127.040501} {\bibfield  {journal}
  {\bibinfo  {journal} {Phys. Rev. Lett.}\ }\textbf {\bibinfo {volume} {127}},\
  \bibinfo {pages} {040501} (\bibinfo {year} {2021})}\BibitemShut {NoStop}%
\bibitem [{\citenamefont {Knizia}\ and\ \citenamefont
  {Chan}(2012)}]{knizia2012density}%
  \BibitemOpen
  \bibfield  {author} {\bibinfo {author} {\bibfnamefont {G.}~\bibnamefont
  {Knizia}}\ and\ \bibinfo {author} {\bibfnamefont {G.~K.-L.}\ \bibnamefont
  {Chan}},\ }\href {https://doi.org/10.1103/PhysRevLett.109.186404} {\bibfield
  {journal} {\bibinfo  {journal} {Phys. Rev. Lett.}\ }\textbf {\bibinfo
  {volume} {109}},\ \bibinfo {pages} {186404} (\bibinfo {year}
  {2012})}\BibitemShut {NoStop}%
\bibitem [{\citenamefont {Rubin}(2016)}]{rubin2016hybrid}%
  \BibitemOpen
  \bibfield  {author} {\bibinfo {author} {\bibfnamefont {N.~C.}\ \bibnamefont
  {Rubin}},\ }\href@noop {} {\bibinfo {title} {A hybrid classical/quantum
  approach for large-scale studies of quantum systems with density matrix
  embedding theory}} (\bibinfo {year} {2016}),\ \Eprint
  {https://arxiv.org/abs/1610.06910} {arXiv:1610.06910} \BibitemShut {NoStop}%
\bibitem [{\citenamefont {Kawashima}\ \emph {et~al.}(2021)\citenamefont
  {Kawashima}, \citenamefont {Coons}, \citenamefont {Nam}, \citenamefont
  {Lloyd}, \citenamefont {Matsuura}, \citenamefont {Garza}, \citenamefont
  {Johri}, \citenamefont {Huntington}, \citenamefont {Senicourt}, \citenamefont
  {Maksymov} \emph {et~al.}}]{kawashima2021efficient}%
  \BibitemOpen
  \bibfield  {author} {\bibinfo {author} {\bibfnamefont {Y.}~\bibnamefont
  {Kawashima}}, \bibinfo {author} {\bibfnamefont {M.~P.}\ \bibnamefont
  {Coons}}, \bibinfo {author} {\bibfnamefont {Y.}~\bibnamefont {Nam}}, \bibinfo
  {author} {\bibfnamefont {E.}~\bibnamefont {Lloyd}}, \bibinfo {author}
  {\bibfnamefont {S.}~\bibnamefont {Matsuura}}, \bibinfo {author}
  {\bibfnamefont {A.~J.}\ \bibnamefont {Garza}}, \bibinfo {author}
  {\bibfnamefont {S.}~\bibnamefont {Johri}}, \bibinfo {author} {\bibfnamefont
  {L.}~\bibnamefont {Huntington}}, \bibinfo {author} {\bibfnamefont
  {V.}~\bibnamefont {Senicourt}}, \bibinfo {author} {\bibfnamefont {A.~O.}\
  \bibnamefont {Maksymov}}, \emph {et~al.},\ }\href@noop {} {\bibinfo {title}
  {Efficient and accurate electronic structure simulation demonstrated on a
  trapped-ion quantum computer}} (\bibinfo {year} {2021}),\ \Eprint
  {https://arxiv.org/abs/2102.07045} {arXiv:2102.07045} \BibitemShut {NoStop}%
\bibitem [{\citenamefont {Mineh}\ and\ \citenamefont
  {Montanaro}(2021)}]{mineh2021solving}%
  \BibitemOpen
  \bibfield  {author} {\bibinfo {author} {\bibfnamefont {L.}~\bibnamefont
  {Mineh}}\ and\ \bibinfo {author} {\bibfnamefont {A.}~\bibnamefont
  {Montanaro}},\ }\href@noop {} {\bibinfo {title} {Solving the hubbard model
  using density matrix embedding theory and the variational quantum
  eigensolver}} (\bibinfo {year} {2021}),\ \Eprint
  {https://arxiv.org/abs/2108.08611} {arXiv:2108.08611} \BibitemShut {NoStop}%
\bibitem [{\citenamefont {Li}\ \emph {et~al.}(2021)\citenamefont {Li},
  \citenamefont {Huang}, \citenamefont {Cao}, \citenamefont {Huang},
  \citenamefont {Shuai}, \citenamefont {Sun}, \citenamefont {Sun},
  \citenamefont {Yuan},\ and\ \citenamefont {Lv}}]{li2021practical}%
  \BibitemOpen
  \bibfield  {author} {\bibinfo {author} {\bibfnamefont {W.}~\bibnamefont
  {Li}}, \bibinfo {author} {\bibfnamefont {Z.}~\bibnamefont {Huang}}, \bibinfo
  {author} {\bibfnamefont {C.}~\bibnamefont {Cao}}, \bibinfo {author}
  {\bibfnamefont {Y.}~\bibnamefont {Huang}}, \bibinfo {author} {\bibfnamefont
  {Z.}~\bibnamefont {Shuai}}, \bibinfo {author} {\bibfnamefont
  {X.}~\bibnamefont {Sun}}, \bibinfo {author} {\bibfnamefont {J.}~\bibnamefont
  {Sun}}, \bibinfo {author} {\bibfnamefont {X.}~\bibnamefont {Yuan}},\ and\
  \bibinfo {author} {\bibfnamefont {D.}~\bibnamefont {Lv}},\ }\href@noop {}
  {\bibinfo {title} {Toward practical quantum embedding simulation of realistic
  chemical systems on near-term quantum computers}} (\bibinfo {year} {2021}),\
  \Eprint {https://arxiv.org/abs/2109.08062} {arXiv:2109.08062} \BibitemShut
  {NoStop}%
\bibitem [{\citenamefont {Kotliar}\ \emph {et~al.}(2006)\citenamefont
  {Kotliar}, \citenamefont {Savrasov}, \citenamefont {Haule}, \citenamefont
  {Oudovenko}, \citenamefont {Parcollet},\ and\ \citenamefont
  {Marianetti}}]{kotliar2006electronic}%
  \BibitemOpen
  \bibfield  {author} {\bibinfo {author} {\bibfnamefont {G.}~\bibnamefont
  {Kotliar}}, \bibinfo {author} {\bibfnamefont {S.~Y.}\ \bibnamefont
  {Savrasov}}, \bibinfo {author} {\bibfnamefont {K.}~\bibnamefont {Haule}},
  \bibinfo {author} {\bibfnamefont {V.~S.}\ \bibnamefont {Oudovenko}}, \bibinfo
  {author} {\bibfnamefont {O.}~\bibnamefont {Parcollet}},\ and\ \bibinfo
  {author} {\bibfnamefont {C.~A.}\ \bibnamefont {Marianetti}},\ }\href
  {https://doi.org/10.1103/RevModPhys.78.865} {\bibfield  {journal} {\bibinfo
  {journal} {Rev. Mod. Phys.}\ }\textbf {\bibinfo {volume} {78}},\ \bibinfo
  {pages} {865} (\bibinfo {year} {2006})}\BibitemShut {NoStop}%
\bibitem [{\citenamefont {Bauer}\ \emph {et~al.}(2016)\citenamefont {Bauer},
  \citenamefont {Wecker}, \citenamefont {Millis}, \citenamefont {Hastings},\
  and\ \citenamefont {Troyer}}]{bauer2016hybrid}%
  \BibitemOpen
  \bibfield  {author} {\bibinfo {author} {\bibfnamefont {B.}~\bibnamefont
  {Bauer}}, \bibinfo {author} {\bibfnamefont {D.}~\bibnamefont {Wecker}},
  \bibinfo {author} {\bibfnamefont {A.~J.}\ \bibnamefont {Millis}}, \bibinfo
  {author} {\bibfnamefont {M.~B.}\ \bibnamefont {Hastings}},\ and\ \bibinfo
  {author} {\bibfnamefont {M.}~\bibnamefont {Troyer}},\ }\href
  {https://doi.org/10.1103/PhysRevX.6.031045} {\bibfield  {journal} {\bibinfo
  {journal} {Phys. Rev. X}\ }\textbf {\bibinfo {volume} {6}},\ \bibinfo {pages}
  {031045} (\bibinfo {year} {2016})}\BibitemShut {NoStop}%
\bibitem [{\citenamefont {Rungger}\ \emph {et~al.}(2019)\citenamefont
  {Rungger}, \citenamefont {Fitzpatrick}, \citenamefont {Chen}, \citenamefont
  {Alderete}, \citenamefont {Apel}, \citenamefont {Cowtan}, \citenamefont
  {Patterson}, \citenamefont {Ramo}, \citenamefont {Zhu}, \citenamefont
  {Nguyen} \emph {et~al.}}]{rungger2019dynamical}%
  \BibitemOpen
  \bibfield  {author} {\bibinfo {author} {\bibfnamefont {I.}~\bibnamefont
  {Rungger}}, \bibinfo {author} {\bibfnamefont {N.}~\bibnamefont
  {Fitzpatrick}}, \bibinfo {author} {\bibfnamefont {H.}~\bibnamefont {Chen}},
  \bibinfo {author} {\bibfnamefont {C.}~\bibnamefont {Alderete}}, \bibinfo
  {author} {\bibfnamefont {H.}~\bibnamefont {Apel}}, \bibinfo {author}
  {\bibfnamefont {A.}~\bibnamefont {Cowtan}}, \bibinfo {author} {\bibfnamefont
  {A.}~\bibnamefont {Patterson}}, \bibinfo {author} {\bibfnamefont {D.~M.}\
  \bibnamefont {Ramo}}, \bibinfo {author} {\bibfnamefont {Y.}~\bibnamefont
  {Zhu}}, \bibinfo {author} {\bibfnamefont {N.~H.}\ \bibnamefont {Nguyen}},
  \emph {et~al.},\ }\href@noop {} {\bibinfo {title} {Dynamical mean field
  theory algorithm and experiment on quantum computers}} (\bibinfo {year}
  {2019}),\ \Eprint {https://arxiv.org/abs/1910.04735} {arXiv:1910.04735}
  \BibitemShut {NoStop}%
\bibitem [{\citenamefont {Cotton}(2003)}]{cotton2003pointgroup}%
  \BibitemOpen
  \bibfield  {author} {\bibinfo {author} {\bibfnamefont {F.~A.}\ \bibnamefont
  {Cotton}},\ }\href@noop {} {\emph {\bibinfo {title} {Chemical applications of
  group theory}}}\ (\bibinfo  {publisher} {John Wiley \& Sons},\ \bibinfo
  {year} {2003})\BibitemShut {NoStop}%
\bibitem [{\citenamefont {Mirman}(1999)}]{pointgroupbook2}%
  \BibitemOpen
  \bibfield  {author} {\bibinfo {author} {\bibfnamefont {R.}~\bibnamefont
  {Mirman}},\ }\href {https://doi.org/10.1142/3994} {\emph {\bibinfo {title}
  {Point Groups, Space Groups, Crystals, Molecules}}}\ (\bibinfo  {publisher}
  {WORLD SCIENTIFIC},\ \bibinfo {year} {1999})\BibitemShut {NoStop}%
\bibitem [{\citenamefont {Butler}(2012)}]{pointgroupbook3}%
  \BibitemOpen
  \bibfield  {author} {\bibinfo {author} {\bibfnamefont {P.~H.}\ \bibnamefont
  {Butler}},\ }\href {https://doi.org/10.1007/978-1-4613-3141-4} {\emph
  {\bibinfo {title} {Point group symmetry applications: methods and tables}}}\
  (\bibinfo  {publisher} {Springer},\ \bibinfo {year} {2012})\BibitemShut
  {NoStop}%
\bibitem [{\citenamefont {Taube}\ and\ \citenamefont
  {Bartlett}(2006)}]{Taube2006}%
  \BibitemOpen
  \bibfield  {author} {\bibinfo {author} {\bibfnamefont {A.~G.}\ \bibnamefont
  {Taube}}\ and\ \bibinfo {author} {\bibfnamefont {R.~J.}\ \bibnamefont
  {Bartlett}},\ }\href {https://doi.org/https://doi.org/10.1002/qua.21198}
  {\bibfield  {journal} {\bibinfo  {journal} {Int. J. Quantum Chem.}\ }\textbf
  {\bibinfo {volume} {106}},\ \bibinfo {pages} {3393} (\bibinfo {year}
  {2006})}\BibitemShut {NoStop}%
\bibitem [{\citenamefont {Nielsen}\ and\ \citenamefont
  {Chuang}(2002)}]{nielsen2002quantum}%
  \BibitemOpen
  \bibfield  {author} {\bibinfo {author} {\bibfnamefont {M.~A.}\ \bibnamefont
  {Nielsen}}\ and\ \bibinfo {author} {\bibfnamefont {I.}~\bibnamefont
  {Chuang}},\ }\href {https://doi.org/10.1119/1.1463744} {\bibfield  {journal}
  {\bibinfo  {journal} {American Journal of Physics}\ }\textbf {\bibinfo
  {volume} {70}},\ \bibinfo {pages} {558} (\bibinfo {year} {2002})}\BibitemShut
  {NoStop}%
\bibitem [{\citenamefont {ANIS}\ \emph {et~al.}(2021)\citenamefont {ANIS},
  \citenamefont {Abraham} \emph {et~al.}}]{Qiskit}%
  \BibitemOpen
  \bibfield  {author} {\bibinfo {author} {\bibfnamefont {M.~S.}\ \bibnamefont
  {ANIS}}, \bibinfo {author} {\bibfnamefont {H.}~\bibnamefont {Abraham}}, \emph
  {et~al.},\ }\href {https://doi.org/10.5281/zenodo.2573505} {\bibinfo {title}
  {Qiskit: An open-source framework for quantum computing}} (\bibinfo {year}
  {2021})\BibitemShut {NoStop}%
\bibitem [{\citenamefont {LaRose}\ \emph {et~al.}(2021)\citenamefont {LaRose},
  \citenamefont {Mari}, \citenamefont {Kaiser}, \citenamefont {Karalekas},
  \citenamefont {Alves}, \citenamefont {Czarnik}, \citenamefont {Mandouh},
  \citenamefont {Gordon}, \citenamefont {Hindy}, \citenamefont {Robertson},
  \citenamefont {Thakre}, \citenamefont {Shammah},\ and\ \citenamefont
  {Zeng}}]{Mitiq}%
  \BibitemOpen
  \bibfield  {author} {\bibinfo {author} {\bibfnamefont {R.}~\bibnamefont
  {LaRose}}, \bibinfo {author} {\bibfnamefont {A.}~\bibnamefont {Mari}},
  \bibinfo {author} {\bibfnamefont {S.}~\bibnamefont {Kaiser}}, \bibinfo
  {author} {\bibfnamefont {P.~J.}\ \bibnamefont {Karalekas}}, \bibinfo {author}
  {\bibfnamefont {A.~A.}\ \bibnamefont {Alves}}, \bibinfo {author}
  {\bibfnamefont {P.}~\bibnamefont {Czarnik}}, \bibinfo {author} {\bibfnamefont
  {M.~E.}\ \bibnamefont {Mandouh}}, \bibinfo {author} {\bibfnamefont {M.~H.}\
  \bibnamefont {Gordon}}, \bibinfo {author} {\bibfnamefont {Y.}~\bibnamefont
  {Hindy}}, \bibinfo {author} {\bibfnamefont {A.}~\bibnamefont {Robertson}},
  \bibinfo {author} {\bibfnamefont {P.}~\bibnamefont {Thakre}}, \bibinfo
  {author} {\bibfnamefont {N.}~\bibnamefont {Shammah}},\ and\ \bibinfo {author}
  {\bibfnamefont {W.~J.}\ \bibnamefont {Zeng}},\ }\href@noop {} {\bibinfo
  {title} {Mitiq: A software package for error mitigation on noisy quantum
  computers}} (\bibinfo {year} {2021}),\ \Eprint
  {https://arxiv.org/abs/2009.04417} {arXiv:2009.04417 [quant-ph]} \BibitemShut
  {NoStop}%
\end{thebibliography}%

\section*{Appendix}
\label{appendix}

\subsection{Point group}
\label{appendix-pointgroup}
In this section we briefly introduce the relevant knowledge of point groups. More detailed and systematic introduction of point groups please refer to~\cite{cotton2003pointgroup,pointgroupbook2,pointgroupbook3}.
The point group is a set of symmetry operations under which the object is indistinguishable from the original geometry. In this paper, we mainly refer the object as a molecule. The so-called point group comes from the origin point being unchanged with arbitrary symmetry operation since all the symmetry elements intersect at it.

\begin{figure}[b]
\begin{center}
\includegraphics[width=\linewidth]{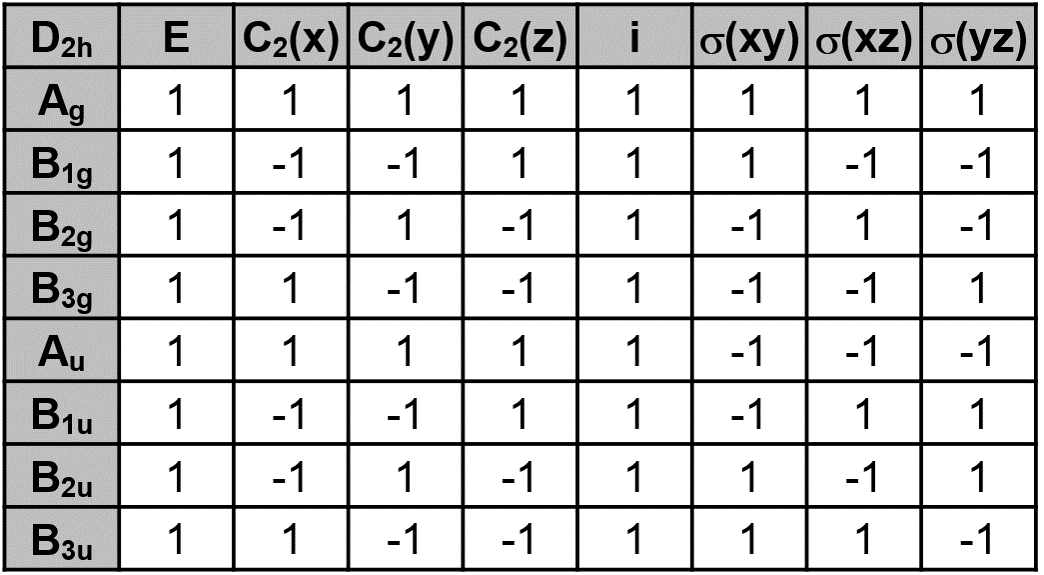}
\caption{\label{characterTable} The character table of $\pg{D_{2h}}$ group for $\mol{BeH_2}$.}
\end{center}
\end{figure}
The symmetry operations include reflection, inversion, rotation and identity operation, which correspond to symmetry elements of mirror planes $\sigma$, inversion center $i$, rotation axes $C_n$ and identities $E$, respectively. Molecules belong to a symmetry point group if it is unchanged under all the symmetry operations of this group.

\begin{figure}[htb]
\begin{center}
\includegraphics[width=\linewidth]{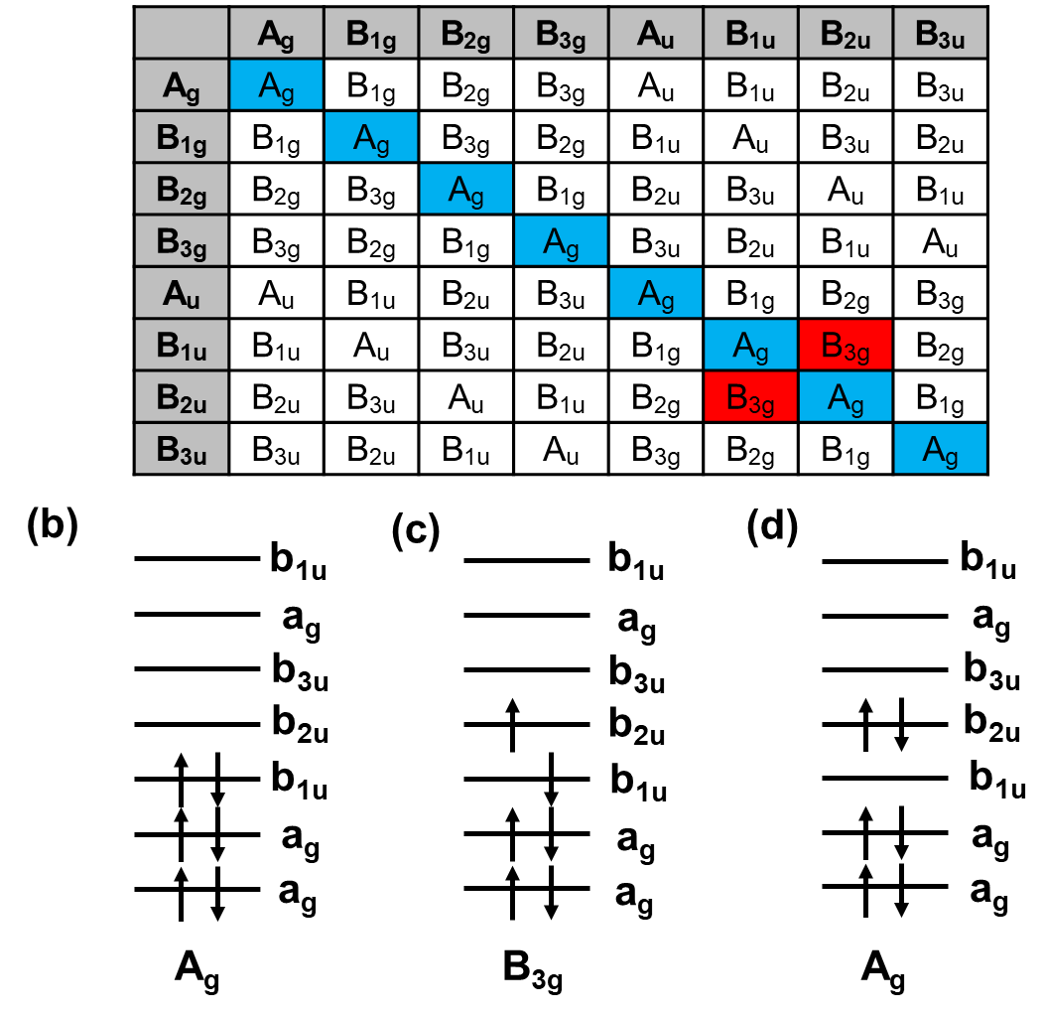}
\caption{\label{symmetry_case} {\bf The irrep of the $\pg{D_{2h}}$ point group for a $\mol{BeH_2}$ molecule.} (a) The product table for the irreducible representations of the $D_{2h}$ group. (b) The electron configuration diagram of $\mol{BeH_2}$ of the reference state, i.e. the Hartree-Fock state with irrep $\pg{A_g}$.
(c) An example of coupled-cluster terms that corresponding to a configuration with different irrep $\pg{B_3g}$ compare to the reference state. This term is a single excitation concerning the reference term. 
(d) An example of coupled-cluster terms that corresponding to a configuration with the same irrep $\pg{A_g}$ as the reference state. 
Note that it is conventionally to label the irrep of the molecular orbital in lower case like $\pg{a_g}$ and the irrep of the molecular state in upper case like $\pg{A_g}$.
}
\end{center}
\end{figure}

In application, the character table and the product table are the essential and frequently used devices for the point group.
Table.~\ref{characterTable} is a typical character table of $\pg{D_{2h}}$ group, in which the rows are the irreducible group representations and the columns are the conjugacy class of the symmetric operations. The table entries are the characters (trace of the matrix) of the symmetric operations under the different irreducible representations. 
Fig.~\ref{symmetry_case}a is the product table of $\pg{D_{2h}}$ group, which exhibits how the character changes under the direct product of the representations. Since the characters of the representation of a direct product are equal to the products of the characters of the representations based on the individual sets of functions, the product table of the point group can be generated by the products of the characters.

Taking the direct product of $B_{1g}$ and $B_{2g}$ as an example, the characters of $B_{1g}$ are {1, -1, -1, 1, 1, 1, -1, -1} and $B_{2g}$ are {1, -1, 1, -1, 1, -1, 1, -1}. The products of the characters under each conjugacy class of the group element are {1, 1, -1, -1, 1, -1, -1, 1}, respectively, which correspond to the characters of $\pg{B_{3g}}$ irreducible representation. So the direct product of $\pg{B_{1g}}$ and $\pg{B_{2g}}$ leads to $\pg{B_{3g}}$.
Similarly, a product table of $\pg{D_{2h}}$ group is constructed according to the result of the direct product between two arbitrary irreps.

\begin{figure}[htb]
\begin{center}
\includegraphics[width=\linewidth]{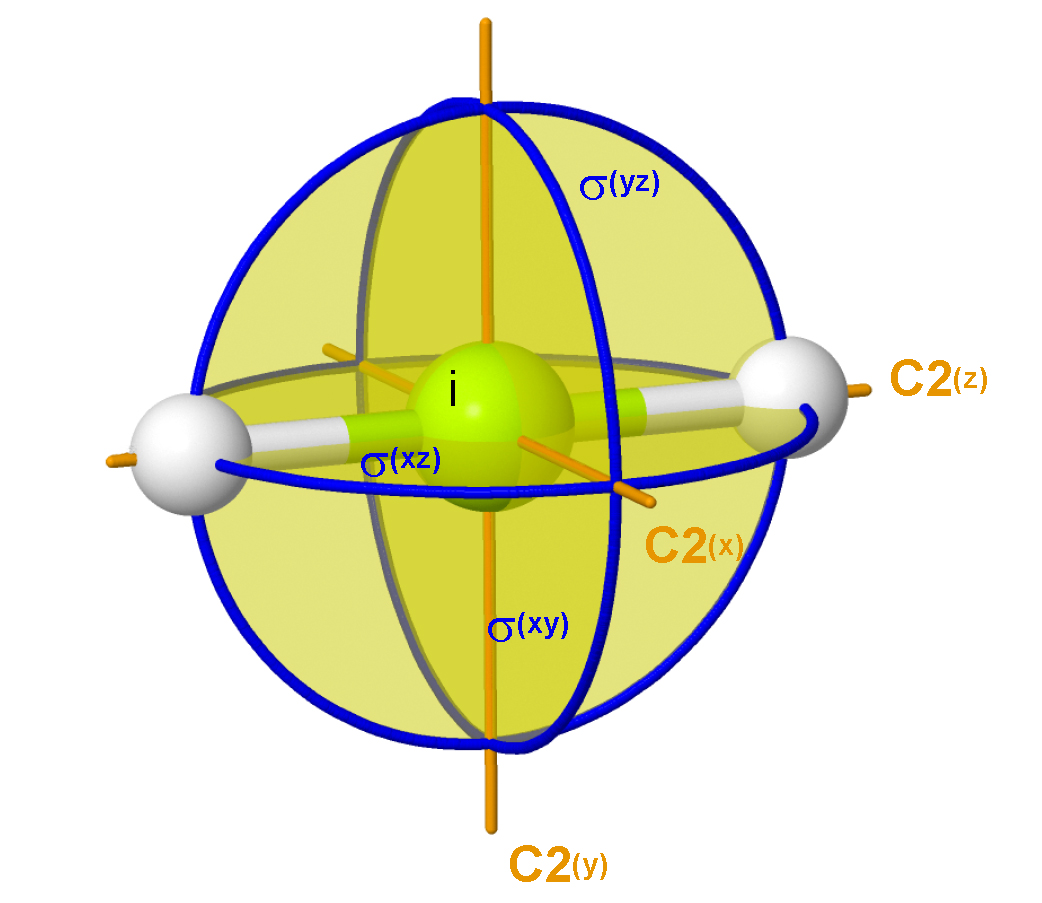} 
\caption{\label{BeH2symelements} The symmetry elements of $\mol{BeH_2}$ in $\pg{D_{2h}}$ point group. Rotation axes $\pg{C_{2}}$, reflection mirrors $\pg{\sigma}$ and inversion center i are depicted in the figure. }
\end{center}
\end{figure}

Here we take $\mol{BeH_2}$ as an example to illustrate detailed steps for our algorithm. $\mol{BeH_2}$ structure belongs to the $\pg{D_{{\infty}h}}$ group, which is nonabelian. In this paper, we only discuss the abelian situation which is easy to handle. Here, we take the $\pg{D_{2h}}$, an Abelian subgroup of $\pg{D_{{\infty}h}}$ with the highest rank, to reduce the parameters of $\mol{BeH_2}$. 
The Cartesian coordinate system is set up with the origin at Be atom and the z-axis along Be-H as shown in Fig.~\ref{BeH2symelements}. 

There are eight symmetry operations corresponding to the same number of symmetry elements for $\mol{BeH_2}$ under $\pg{D_{2h}}$ group. Three $\pg{C_{2}}$ rotation axes are along the x-, y-, z-axis, and three mirrors are in the xy-, yz-, zx- planes which are perpendicular with each other and intersect in a $\pg{C_{2}}$ rotation axis. The rest of the symmetry elements are the inversion center at the place of Be atom and the identity.

\begin{figure}[tb]
\begin{center}
\includegraphics[width=\linewidth]{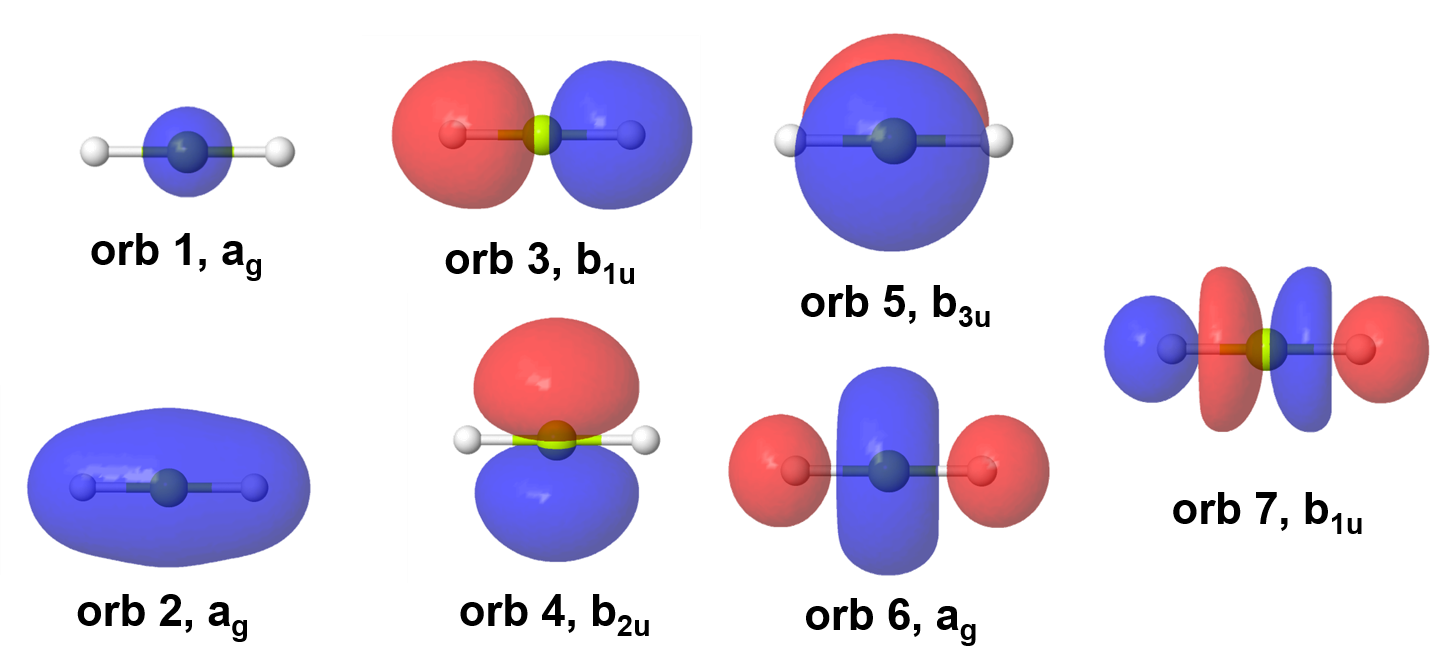}
\caption{\label{MOsurfaceBeH2} The isosurface of the wavefunction of molecular orbital (isovalue is 0.03 a.u.).}
\end{center}
\end{figure}
We display molecular orbitals of $\mol{BeH_2}$ to illustrate how to understand the spatial symmetry in it combining with the character table. 
The isosurface of molecular orbitals are depicted in Fig.~\ref{MOsurfaceBeH2}, in which the blue surface represents the positive value, and the red surface represents the negative one. The character in character Table.~\ref{characterTable} indicates how the sign of the certain irreducible representation changes under the particular symmetry operation. For the $\pg{A_g}$ orbitals, the sign of the wavefunction maintains under arbitrary symmetry operation. Thus, all the characters are 1. For the $\pg{B_{1u}}$ orbitals like orb 3 and orb 7, the sign of the wavefunction is kept after rotated along z-axis or reflected by xz-plane and yz-plane. But the sign exchanges after rotated by x-axis and y-axis or reflected by xy-plane. Thus, the characters are -1 for i, $\pg{C_2(x)}$, $\pg{C_2(y)}$ and $\pg{\sigma(xy)}$ while +1 for the other operations.
The product table is then generated according to the character table as shown in Fig.~\ref{symmetry_case}(a). Then it is possible to get the irrep of the excited term.
The key step is to determine the irrep of the excited terms (or the excitation operators) and retain only the appropriate items by referring to the product table. The wavefunction of the reference state is expressed as the Slater-determinant of a collection of molecular orbitals, i.e. $\left|\Psi_{0}\right\rangle=\left|\phi_{1}\overline{\phi}_{1}\phi_{2}\overline{\phi}_{2}...\phi_{n}\overline{\phi}_{n}\right\rangle$, where $\phi$ is the occupied molecular spin-orbital and the bar indicates the different spin. The irrep of the molecular state is determined from the direct product of the molecular spin-orbitals.

Again we go back to $\mol{BeH_2}$ in $\pg{D_{2h}}$ point group.
In Fig.~\ref{symmetry_case}(a) we provide a product table for $\pg{D_{2h}}$ group, where the product relationship between two irrep in $\pg{D_{2h}}$ is given. Fig.~\ref{symmetry_case}(b-d) presents three different electronic configuration diagrams of $\mol{BeH_2}$. In Fig.~\ref{symmetry_case}(b) we present the Hartree-Fock ground state, i.e. the reference state, in which three lowest orbitals with the irrep of $\rm{a_{g}}$, $\rm{b_{1u}}$ and $\rm{a_{g}}$ are doubly occupied. By referring to the product table in Fig.~\ref{symmetry_case}(a), the irrep of the reference state is easily obtained, as $(a_g{\otimes}a_g){\otimes}(b_{1u}{\otimes}b_{1u}){\otimes}(a_g{\otimes}a_g)=a_g{\otimes}a_g{\otimes}a_g=A_g$. Usually, the irrep of the molecular wavefunction is written in uppercase to distinguish from the irrep of the molecular orbitals.
In our method, after determining the irrep of the reference state, then all the possible single Slater-determinant excited terms $\hat{t}\left|\Psi_{0}\right\rangle$ are traversed to check whether their share the same irrep as the reference term. We present two specific examples in Fig.~\ref{symmetry_case}(c,d). For the single excitation term $\hat{t}^4_3\left|\Psi_{0}\right\rangle$ in Fig.~\ref{symmetry_case}(c), one electron is excited from the 3rd orbital to the 4th orbital. Its corresponding irrep is $(a_g{\otimes}a_g){\otimes}(a_{g}{\otimes}a_{g}){\otimes}(b_{1u}){\otimes}(b_{2u})=a_g{\otimes}a_g{\otimes}b_{1u}{\otimes}b_{2u}=B_{3g} \not= A_g$, which is not expected, and the exciting operator $\hat{t}^4_3$ should be excluded from constructing the cluster operator. As shown in Fig.~\ref{symmetry_case}(d), both two electrons at the 3rd orbital are excited from to the 4th orbital. The irrep is similarly evaluated as $(a_g{\otimes}a_g){\otimes}(b_{1u}{\otimes}b_{1u}){\otimes}(b_{2u}{\otimes}b_{2u})=a_g{\otimes}a_g{\otimes}a_g=A_g$, which is the same as the reference term. So the exciting operator $\hat{t}^{44}_{33}$ is remained in constructing the cluster operator.

In total, there are 12 ($3\times4$) single excited states
and 78 
double excited states for $\mol{BeH_2}$. Only 23 excitations shared the same irreducible representation $\pg{A_g}$ as the reference state. Thus, only the certain 23 excitation operators are included when construct the ansatz in the following VQE steps.

\subsection{Validation of the Point Group Symmetry Reduction in UCC}
\label{Proof}
Here we give a short derivation to valid the algorithm. 
In the UCC theory, the wavefunction of a chemical system $|\Psi\rangle$ is constructed from the reference wavefunction $\left|\Psi_{0}\right\rangle$, which is usually a Hartree-Fock Slater determinant, by applying $e^{\hat{T}-\hat{T}^{\dagger}}$ as
\begin{equation}
|\Psi\rangle=e^{\hat{T}-\hat{T}^{\dagger}}|\Psi_0\rangle,
\end{equation} 
and the Schr\"odinger equation can be written as
\begin{equation}\hat{H}|\Psi\rangle=E_{ucc}|\Psi\rangle.\end{equation}

For ensuring $|\Psi\rangle$ to be a solution of the Schr\"odinger equation, the cluster operator $e^{\hat{T}-\hat{T}^{\dagger}}$ here is not arbitrary but has to satisfy some conditions. In the following we will present how to reduce the number of operators and terms in the unitary coupled-cluster wavefunction $|\Psi\rangle$. Note that the wavefunction here refers to a solution of the Schr\"odinger equation instead of a quantum state prepared on a quantum computer.

Here we have assumed that both $\left|\Psi_{0}\right\rangle$ and $|\Psi\rangle$ are non-degenerated states. For a symmetry operation $\hat{R}_{i}$ belongs to an Abelian point group G, the Hamiltonian is commute with it, $\hat{H} \hat{R}_{i}=\hat{R}_{i} \hat{H}$.
Thus we have
\begin{equation}
\hat{H} \hat{R}_{i}|\Psi\rangle=\hat{R}_{i} \hat{H}|\Psi\rangle=E_{ucc} \hat{R}_{i}|\Psi\rangle=E_{ucc}\left(\hat{R}_{i}|\Psi\rangle\right)
\end{equation}
It indicates that $\hat{R}_{i}|\Psi\rangle$ is also the eigenstate of the Hamiltonian. So we come to
\begin{equation}
\exists \gamma: \hat{R}_{i}|\Psi\rangle=\gamma|\Psi\rangle
\end{equation}
where the value of $\gamma$ is the irreducible character that can be looked up from the character table of the corresponding point group. The irreducible characters in the Abelian point group are always 1 or -1.

In the following, we will show that $\left|\Psi_{0}\right\rangle$ and $|\Psi\rangle$ belong to the same irrep:
\begin{equation}
D\left(\left|\Psi_{0}\right\rangle\right)=D(|\Psi\rangle)
\end{equation}
where $D$ is the irrep of the corresponding wavefunction.

By applying the symmetry operator to $\left|\Psi_{0}\right\rangle$ and $|\Psi\rangle$ respectively, we have
\begin{equation}
\left\langle\Psi_{0} \mid \Psi\right\rangle=\left\langle\Psi_{0}\left|R^{\dagger} R\right| \Psi\right\rangle=c_{0} c_{1}\left\langle\Psi_{0} \mid \Psi\right\rangle
\end{equation}
We note that a reasonable solution $|\Psi\rangle$ perturbed from  $\left|\Psi_{0}\right\rangle$ should be overlapping with the Hartree-Fock determinant $\left|\Psi_{0}\right\rangle$ \cite{Taube2006}
\begin{equation}
\left\langle\Psi_{0} \mid \Psi\right\rangle \neq 0
\end{equation}

It indicates that
\begin{equation}
c_{0} c = 1
\end{equation}
In an Abelian point group, the characters are either 1 or -1. So we know
\begin{equation}
c_{0}=c,
\end{equation}
It means that 
\begin{equation}
\forall \hat{R}_{i} \in G: \hat{R}_{i}|\Psi_{0}\rangle=c_{i}\left|\Psi_{0}\right\rangle, \hat{R}_{i}|\Psi\rangle=c_{i}|\Psi\rangle.
\end{equation}

This concludes that each symmetry operation $\hat{R}_i \in G $ acting on $|\Psi\rangle$ and $\left|\Psi_{0}\right\rangle$ will lead to the same character, and thus they belong to the same $irrep$.

The cluster operator $e^{\hat{T}-\hat{T}^{\dagger}}$ might be expanded by Taylor expansion,
\begin{equation}
e^{\hat{T}-\hat{T}^{\dagger}}=1+(\hat{T}-\hat{T}^{\dagger})+\frac{1}{2!}(\hat{T}-\hat{T}^{\dagger})+......
\end{equation}

Thus $|\Psi\rangle$ can be written as the linear combination of Slater determinants:
\begin{equation}
|\Psi\rangle=k(\left|\Psi_{0}\right\rangle+\sum_{i, a} c_{i}^{a}\left|\Psi_{i}^{a}\right\rangle+\sum_{i, j, a, b} c_{i j}^{a b}\left|\Psi_{i j}^{a b}\right\rangle+\ldots \ldots),
\end{equation}
where $k$ is the normalized coefficient and $c$ is the summarized coefficient of each excited term including connected and disconnected terms.

As we have presented that the $|\Psi\rangle$ and $\left|\Psi_{0}\right\rangle$ are of the same irrep, so all the terms in the expansion are required to have the same irrep as the reference wavefunction. If $D\left(c_{i}^{a}\left|\Psi_{i}^{a}\right\rangle\right) \neq D\left(\left|\Psi_{0}\right\rangle\right)$, 
$c_{i}^{a}\left|\Psi_{i}^{a}\right\rangle$ must be ZERO. 

It indicates that only the excitation and de-excitation operators, $\hat{t}$ and $\hat{t}^{\dagger}$, belong to the corresponding $irrep$ can survive. Otherwise, the excited term $\left|\Psi_{k}\right\rangle$ belongs to other $irrep$ would appear in the expansion of $|\Psi\rangle$. So in our algorithm, we filter out the excitation (and the de-excitation) operators belong to the different $irrep$:
\begin{equation}
\forall D\left(\hat{t}\left|\Psi_{0}\right\rangle\right) \neq D\left(\left|\Psi_{0}\right\rangle\right) : \hat{t}=0.
\end{equation}
Considering the ansatz after Trotterization, it can alternatively be stated as:
\begin{equation}
\forall D\left(e^{\hat{t}-\hat{t}^{\dagger}}\left|\Psi_{0}\right\rangle\right) \neq D\left(\left|\Psi_{0}\right\rangle\right) : e^{\hat{t}-\hat{t}^{\dagger}}=1.
\end{equation}
When applying this method, it is implicitly assumed that the irrep of HF reference state is of the correct irrep of the true ground state. We will discuss how this assumption might affect the application of the proposed algorithm. For the simple closed shell molecules like $\mol{BeH_2}$~(meaning there is no unpaired electron), the ground state is necessarily to be of the totally symmetric irrep, such as $\pg{A_g}$ in $\pg{D_{2h}}$ group, $\pg{A_1}$ in $\pg{
C_{2v}}$ group and $\pg{A'}$ in $\pg{C_{s}}$ group. This conclusion applies to most of the organic compounds and simple ionic compounds including all the cases listed in this work~\cite{cotton2003pointgroup}. However, when it comes to molecules with unpaired electrons, the irrep of the ground state can not get determined directly. In such situations, the ground state evaluated by Hartree-Fock is a good starting point and usually predicts the correct irrep of the true ground state. For more complex cases where a single Slater determinant is not a good approximation, multiple attempts of the Hartree-Fock ground state and low lying excited states may be needed for reaching the true ground state, as what have been done in conventional quantum chemical calculations.

\subsection{Parameters Counting and Reduction}
The wavefunction $|\Psi\rangle$ constructed from the UCCSD ansatz  $|\Psi\rangle=e^{\hat{T}-\hat{T}^{\dagger}}|\Psi_0\rangle$, is actually implemented by fermion-spin transformations (such as Jordan-Wigner transformation and Bravyi–Kitae transformation) and first-order Trotterization. After that, the quantum circuit can be treated as the product of a series of time evolution of Pauli strings like $U(\Vec{\theta})=e^{i\theta_1 \hat{P}_1}e^{i\theta_2 \hat{P}_2}e^{i\theta_3 \hat{P}_3}...$, where $\hat{P}_i$ are Pauli strings. 

For example, if variational parameter $\theta$ is real, after Jordan-Wigner transformation, the single excitation operators become
\begin{equation}
\begin{aligned}
& \hat{t}_i^{a} - \hat{t}_i^{a\dagger} 
=\theta _{i}^{a} (a_{a}^{\dagger } a_{i} -a_{i}^{\dagger } a_{a} )\\
 & =\frac{\theta _{i}^{a}}{4} [(\sigma _{a}^{x} -i\sigma _{a}^{y} )(-\sigma _{i}^{x} +i\sigma _{i}^{y} )-(\sigma _{i}^{x} -i\sigma _{i}^{y} )(-\sigma _{a}^{x} +i\sigma _{a}^{y} )]\left( \bigotimes _{k=i+1}^{a-1} \sigma _{k}^{z}\right)\\
 & =i\frac{\theta _{i}^{a}}{2} (\sigma _{a}^{y} \sigma _{i}^{x} -\sigma _{a}^{x} \sigma _{i}^{y} )\left( \bigotimes _{k=i+1}^{a-1} \sigma _{k}^{z}\right)
\end{aligned}
\end{equation}
  $\left[ \sigma _{a}^{y} \sigma _{i}^{x} ,\sigma _{m}^{x} \sigma _{n}^{y}\right] =0$ if $m\neq a,n\neq i$. After first-order Trotterization, the unitary operators may looks like $e^{i\frac{\theta _{i}^{a}}{2} \sigma _{a}^{y} \sigma _{k}^{z}... \sigma _{i}^{x}}, e^{i\frac{\theta _{i}^{a}}{2} \sigma _{a}^{x} \sigma _{k}^{z}... \sigma _{i}^{y}}$, etc.
Similarly, the double excitation operators will become
\begin{equation}
\begin{aligned}
& \hat{t}_{ij}^{ab}  - \hat{t}_{ij}^{ab\dagger}
=\theta _{ij}^{ab} (a_{a}^{\dagger } a_{b}^{\dagger } a_{i} a_{j}-a_{j}^{\dagger } a_{i}^{\dagger } a_{b} a_{a})\\
 & =\frac{\theta _{ij}^{ab}}{16} [(\sigma _{a}^{x} -i\sigma _{a}^{y} )(\sigma _{b}^{x} +i\sigma _{b}^{y} )(\sigma _{i}^{x} +i\sigma _{i}^{y} )(-\sigma _{j}^{x} +i\sigma _{j}^{y} )\\
 & \ \ -(\sigma _{a}^{x} +i\sigma _{a}^{y} )(-\sigma _{b}^{x} +i\sigma _{b}^{y} )(\sigma _{i}^{x} -i\sigma _{i}^{y} )(\sigma _{j}^{x} +i\sigma _{j}^{y} )] \\
 & \left( \bigotimes _{k=m+1}^{i-1} \sigma _{k}^{z}\right)\left( \bigotimes _{p=n+1}^{j-1} \sigma _{p}^{z}\right)\\
 & =\frac{\theta _{ij}^{ab}}{8} (i\sigma _{a}^{y} \sigma _{b}^{x} \sigma _{i}^{x} \sigma _{j}^{x} -i\sigma _{a}^{x} \sigma _{b}^{y} \sigma _{i}^{x} \sigma _{j}^{x} -i\sigma _{a}^{x} \sigma _{b}^{x} \sigma _{i}^{y} \sigma _{j}^{x}\\
 &+i\sigma _{a}^{x} \sigma _{b}^{x} \sigma _{i}^{x} \sigma _{j}^{y}-i\sigma _{a}^{y} \sigma _{b}^{y} \sigma _{i}^{y} \sigma _{j}^{x} +i\sigma _{a}^{y} \sigma _{b}^{y} \sigma _{i}^{x} \sigma _{j}^{y} \\
 &+i\sigma _{a}^{y} \sigma _{b}^{x} \sigma _{i}^{y} \sigma _{j}^{y} -i\sigma _{a}^{x} \sigma _{b}^{y} \sigma _{i}^{y} \sigma _{j}^{y} )\left( \bigotimes _{k=m+1}^{i-1} \sigma _{k}^{z}\right)\left( \bigotimes _{p=n+1}^{j-1} \sigma _{p}^{z}\right),
\end{aligned}
\end{equation}
and there will be similar time evolution of Pauli strings after Trotterization.

  To realize the time evolution of Pauli strings, for instance, the operator $e^{i\theta \sigma^z_{1} \sigma^z_{2} \sigma^z_{3} \sigma^z_{4}}$ may be decomposed into single-qubit rotation gates and two-qubit CNOT gate \cite{nielsen2002quantum} as 
  \begin{equation} \label{z1z2z3z4}
    \Qcircuit @C=1em @R=.7em {
	    1& &\qw & \ctrl{1} & \qw & \qw  &\qw  & \qw   & \qw  & \ctrl{1} &\qw &\qw \\
		2& &\qw & \targ & \ctrl{1} & \qw  &\qw  & \qw  & \ctrl{1} & \targ &\qw &\qw \\
		3& &\qw & \qw & \targ & \ctrl{1} & \qw  & \ctrl{1}  & \targ   &\qw &\qw &\qw \\
		4& &\qw & \qw & \qw & \targ & \gate{R_z(2\theta)} & \targ & \qw &\qw &\qw &\qw
		}.
  \end{equation}   
  Other operators may be decomposed to similar circuits other than some single rotations. Accroding to the point group constrain, the symmetry forbidden $e^{\hat{t}-\hat{t}^{\dagger}}$ will be eliminated and the corresponding Pauli strings in the quantum circuit shall disappear, leading to significant reduction to the circuit depth.

\begin{figure}[htb]
\begin{center}
\includegraphics[width=\linewidth]{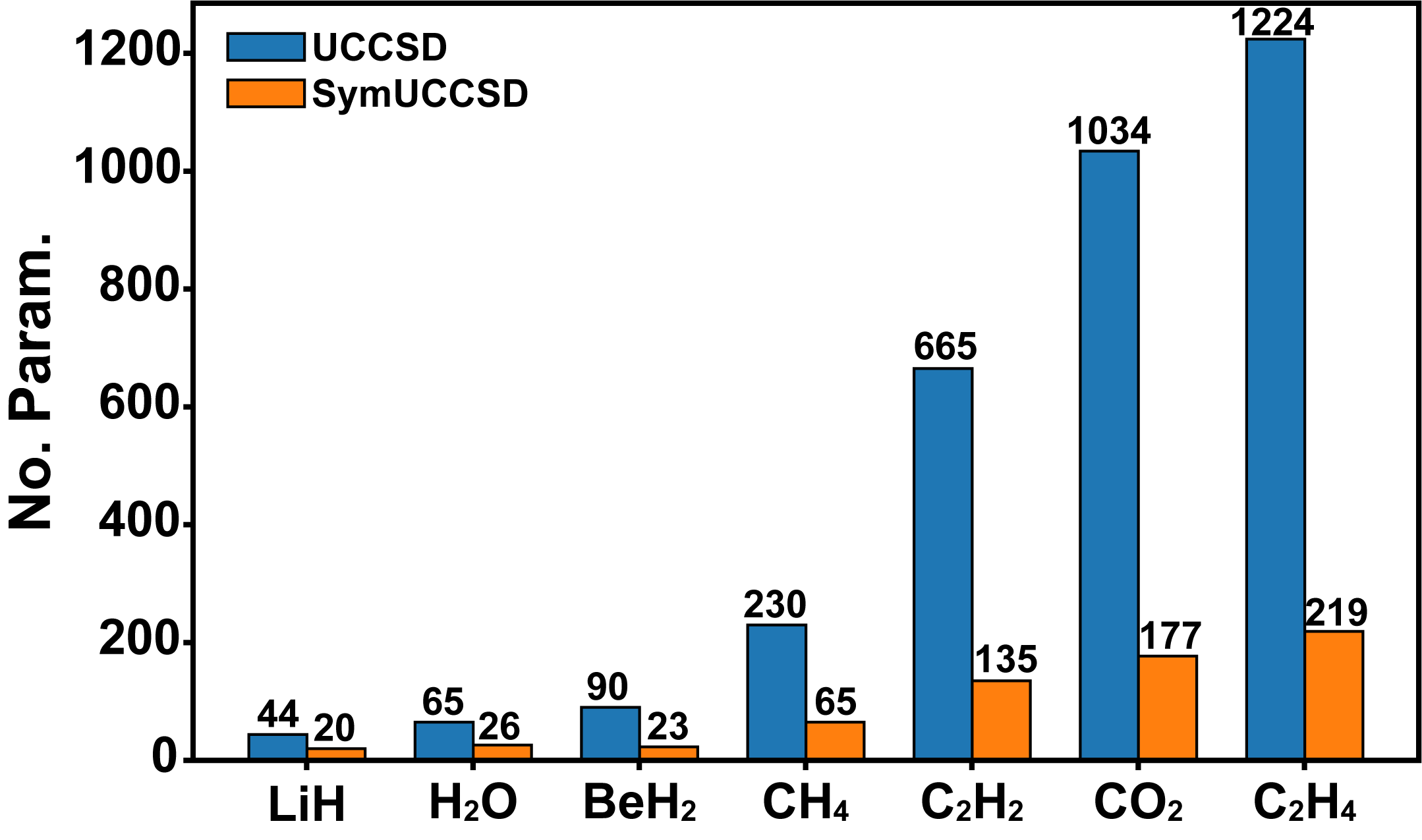}
\caption{\label{parameter_reduction} {\bf The number of parameters for a series of molecules before and after the reduction by point group symmetry.} We observed that the ratio of the necessary parameters decreases slightly as the size of the system increases, ranging from 25.6\% of $\mol{BeH_2}$, to 20.3\% of $\mol{C_2H_2}$ and to $17.9\%$ of $\mol{C_2H_4}$, all of which are of the same $\rm{D_{2h}}$ symmetry, indicating better efficient for larger molecules.}
\end{center}
\end{figure}

In this work, we use the number of parameters (equal to the number of excitation operators) to measure the compactness of ansatz. Considering a restricted closed shell molecule with $n$ occupied spatial orbitals and $m$ virtual spatial orbitals, the number of parameters is count in the following principles:
1. For the single excitation, the number of possible excitation operators are $m \times n$.
2. For the double excitation, the number of possible excitation is computed by the pair excitation $\hat{t}_{ii}^{aa}$ with $m \times n$ terms plus the direct product of the two different single excitations with $C_{mn}^{2}=m \times n(m \times n-1)/2$ terms. 
We apply this method to a series of molecules (without further VQE steps) and compare parameter numbers in ansatz before and after symmetry reduction in Fig.~\ref{parameter_reduction}. The number of the parameters grows much slower after filtering out the symmetry forbidden terms, indicating a shallower quantum circuit and a smaller need of computational resources.
 
\subsection{SymUCCSD Compared and Combined with ADAPT-VQE}
\label{Symm_Adapt_VQE}
The point group employed in SymUCCSD can compare and combine with adapt type ansatz.
One of the famous adapt vqe is ADPAT-VQE proposed in~\cite{adaptvqe}.

The ADPAT-VQE screening operators from a predefined pool to construct the circuit of ansatz, and the operator that will result in larger gradient to the energy would be chosen. In ADAPT-VQE, The operator pool is composed of a set of spin-adapted single and double excitation operators. The screening step is terminated when the average gradient of the operators is smaller than the predefined threshold $\epsilon$. 

In this section, we present the results of ADAPT-VQE~\cite{adaptvqe} as both comparison and combination with SymUCCSD. The calculations are performed using MindQuantum simulator~\cite{mindquantum} with STO-3G basis set~\cite{sto3g}, and the $\epsilon$ is set to $10^{-2}$.

As shown in table~\ref{tab:adaptvqe}, we have performed a series of ADAPT-VQE calculations with different initial operator pools. First, as a comparison in $2^{ed}$ and $7^{th}$ rows of Table~\ref{tab:adaptvqe}), we observe that SymUCCSD reach similar accuracy with slightly more parameters~(23 parameters compare to 18 parameters), when compared with ADAPT-VQE. However, the SymUCCSD does not need to evaluate the gradient of each operator on quantum computers, the evaluation of which will be extra measurement overhead on real quantum devices.

Besides, we also try to combine SymUCCSD with ADAPT-VQE method. As shown in the $2^{ed}$ row of Table~\ref{tab:adaptvqe}, initialize the operator pool of ADAPT-VQE based on the point group symmetry. The number of operators are reduced from 90 to 23 as what has been done in SymUCCSD. By taking this reduced initial operator pool, the number of final operators and energy are the same as those using default full operator pool.

To get more insights of ADAPT-VQE and point group symmetry, we further carried out two ADAPT-VQE calculations as shown in Table~\ref{tab:adaptvqe}) row $4^{th}$ and row $7^{th}$. Both of the calculations fail with the these initial operator pools. For the initial operator pools with randomly selected 23 operators, the algorithm only choose 5 operators in the final, and the final energy is far from the energy calculated from FCI or CCSD. While for the initial operator pool with 67 operators, none of operators survive in ADAPT-VQE process and the HF level energy is obtained. These results indicate that the valid operators in ADAPT-VQE indeed conserve the point group symmetry. 

In summary, the point group symmetry reduction in SymUCCSD can not only efficiently save the computational resource on its own, but also compatible with other methods such as ADAPT-VQE to further reduce quantum resource requirement(here means extra quantum measurement overhead).

\begingroup
\squeezetable
\begin{center}
\begin{table}
\caption {\label{tab:adaptvqe} Numerical simulations of ADAPT-VQE($\epsilon$~= 0.01) and SymUCCSD methods. The size of the initial and final operator pool is listed in the $1^{st}$ and $2^{ed}$ columns. The energy with respect to FCI energy is given in the $4^{th}$ column. The HF, CCSD and FCI energies calculated using PySCF~\cite{PySCF_1,PySCF_2} package and SymUCCSD energy is calculated using the method as mentioned in manuscript.} 

\begin{ruledtabular}
\begin{tabular}{ lllll  }
 Methods & Initial Operators & Final Operators & Energy/Ha. & $\Delta$E/mHa.  \\ \hline
ADAPT-VQE & 90~(Origin) & 18 & -15.5948 & 0.4 \\
ADAPT-VQE & 23~(Same Irrep.) & 18 & -15.5948 & 0.4\\
ADAPT-VQE & 23~(Random) & 5 & -15.5608 &34.4 \\
ADAPT-VQE & 67~(Different Irrep.)& 0 & -15.5603 & 34.9 \\ \hline
UCCSD & 90 & 90 &  -15.5948 & 0.4\\
SymUCCSD & 23 & 23 & -15.5948 & 0.4\\
HF & &  &  -15.5603 & 34.9\\
CCSD & &  & -15.5948 & 0.4\\
FCI &  & & -15.5952& -0-\\

\end{tabular}
\end{ruledtabular}
\end{table}
\end{center}
\endgroup

\subsection{Numerical Simulations with Noise and Measurement Shots}
\label{noise_measurement}

To assess the performance of the SymUCCSD on real quantum device, we evaluate the influence introduced by imperfect device noise and number of measurement shots. We compare SymUCCSD and UCCSD ansatzes for 8-qubit system of $\mol{H_4}$ chain with equal space~(r=1 \AA)~using STO-3G basis set~\cite{sto3g} and carry out the numerical simulation using Qiskit toolkit with QASM simulator with and without noise models~\cite{Qiskit}.

\begin{figure}[htb]
\begin{center}
\includegraphics[width=\linewidth]{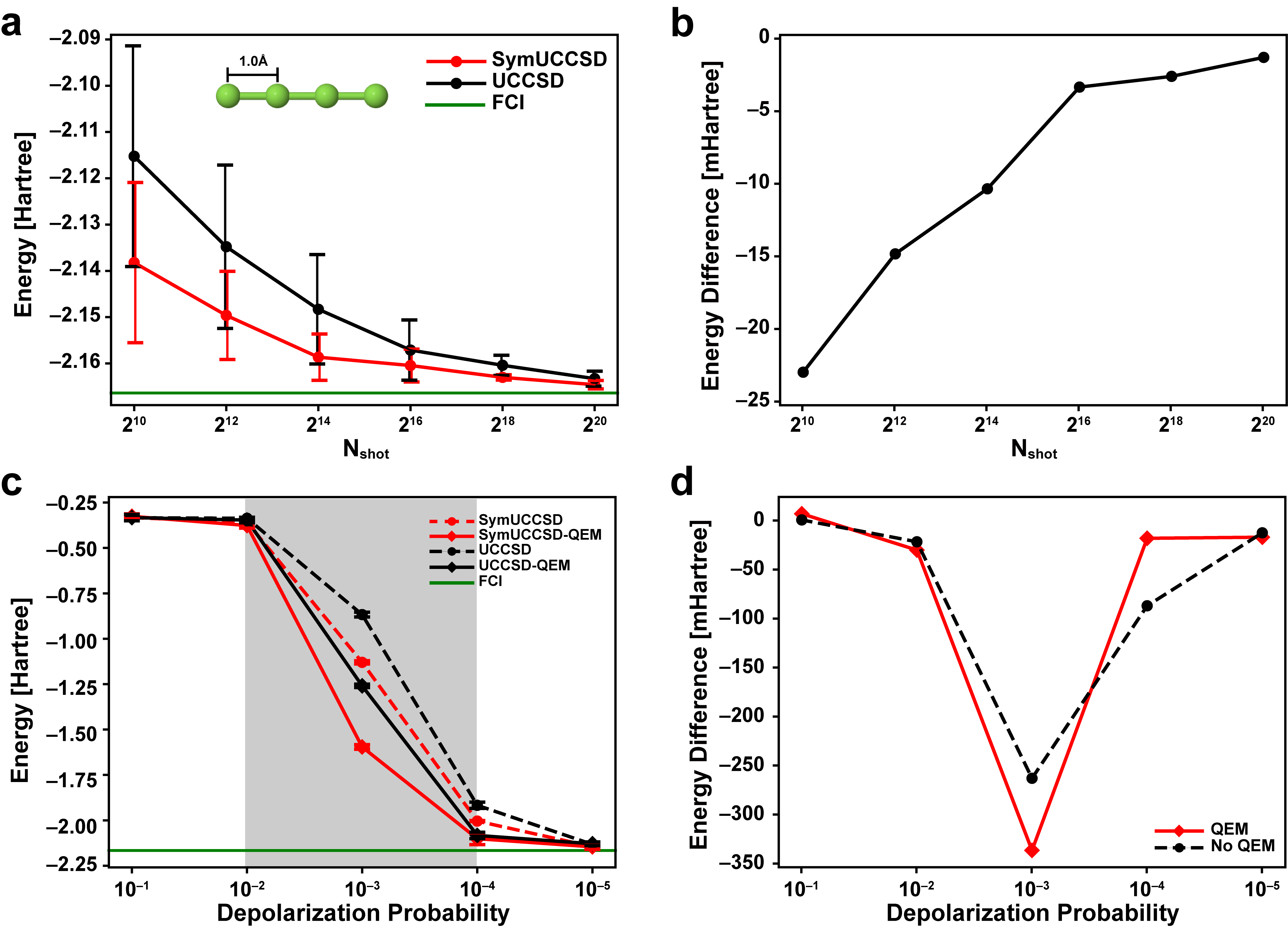}
\caption{\label{noise_simulation} {\bf The numerical simulation of $\bf{H_4}$ molecule with uncertainties.} (a)The energy vs. the number of shots without introducing noise model. (b)The difference between energies calculated by SymUCCSD and UCCSD ansatzes. (c)The energy vs. the depolarization probability. The depolarizing noise model provided by Qiskit~\cite{Qiskit} is used. The quantum error mitigation~(QEM) is performed using zero-noise extrapolation based on a linear fit implemented by mitiq~\cite{Mitiq} package. The magnitude of the standard deviation is tens of mHartree, which makes the error bar look narrow due to the large scaling of y-axis. (d)The difference between energies calculated by SymUCCSD and UCCSD ansatzes.}
\end{center}
\end{figure}

In Fig.~\ref{noise_simulation}(a), we show the VQE energy without any noise. The energy decrease as the increasing of the number of shots. We observed that the energy of SymUCCSD converges faster and have small fluctuation than that of UCCSD as the increasing of measurement shots. This relative fluctuation in UCCSD mainly come from that the very small part of computational-basis state out of the desired irrp. The energy differences between SymUCCSD and UCCSD reach \~ 1mHa at the large number of shots $2^{20}$, as shown in Fig.~\ref{noise_simulation}(b).

In Fig.~\ref{noise_simulation}(c, d). we consider the effect of noise by introducing the depolarizing noise model. The quantum error mitigation (QEM) is performed using zero-noise extrapolation based on a linear fit implemented by mitiq package~\cite{Mitiq}. We observed the SymUCCSD ansatz is more robust to noise~(particularly in the range of $10^{-2}$ to $10^{-3}$, which is the current state of art for real quantum device) in both situations with and without error mitigation, respectively, which is understandable since SymUCCSD has much shallower circuit. 

In a word, with the help of the point group symmetry reduction, we can construct a more compact ansatz SymUCCSD with less parameters. It may not only reduce the computational cost but also provide better accuracy on real quantum devices.

\end{document}